 \documentclass[aps,12pt,floatfix,tightenlines,showkeys,showpacs]{revtex4}
 \usepackage{amsmath, amsthm, amssymb}
 \usepackage{graphicx} 
\usepackage{subfigure}
\newcommand{\be}{\begin{equation}}
\newcommand{\ee}{\end{equation}}
\newcommand{\bea}{\begin{eqnarray}}
\newcommand{\eea}{\end{eqnarray}}
\newcommand{\eq}[1]{Eq.~(\ref{#1})}

\begin {document}

\title{\bf \hskip10cm NT@UW-13-03\\
{Color transparency in     the       reaction\\ $\gamma^*+A\to\rho+p+(A-1)^*$ 
  	}}
                                                                           
\author{Gary T.~Howell, Gerald A.~Miller}
\affiliation{Department of Physics, Univ. of Washington\\
Seattle, WA 98195-1560}
\date{\today}

\begin{abstract}
We  study the nuclear transparency $T$ for the exclusive reaction $\gamma^*+A\to\rho+p+(A-1)^*$ at incident virtual photon energies $\nu \ge$ 10 GeV 
to investigate the separate dependence on the photon virtuality, $Q^2$ and the 4-momentum-transfer-squared to the knocked-out proton, $t$. If the effects of color transparency are included, $T$
 shows significant variation with $t$  even for small values of $Q^2$ for fixed values of the coherence length $l_c$, and also shows significant increase as $Q^2$ is increased at  fixed $l_c$ and $t$. The value of $T$ is found to depend strongly on the phase space over which it is measured. 
\end{abstract}    


\maketitle

\section{Introduction}
\label{sec:intro}

Color Transparency is a prediction of perturbative Quantum Chromodynamics which asserts that when a hadron undergoes a high-momentum-transfer elastic or quasi-elastic reaction inside a nucleus, an incoming or outgoing hadron experiences reduced interactions with the nucleons of the nucleus, compared to their interaction in free-space. See the reviews~\cite{Frankfurt:1992dx,Frankfurt:1994hf,Jain:1995dd,Miller:2007zzd}. 

For example, in the quasielastic scattering of an electron from a nucleus accompanied by proton knockout, $A(e,e'p)(A-1)$, perturbative QCD predicts that if the momentum transfer from the electron to the proton is large enough, the knocked-out proton will experience reduced interactions with the rest of the nucleons on its way out of the nucleus.  For very large momentum transfer, the fast moving proton would not interact with the other nucleons at all.  The quantity that characterizes the initial- and final-state interactions of the projectile and/or outgoing hadrons is called the nuclear transparency, which is defined as the ratio of two cross-sections:
\be
\label{transpardef}
T\equiv \frac{\sigma}{\sigma_{PWIA}}
\ee
where $\sigma$ is the actual measured cross-section for the reaction occuring in a nucleus, and $\sigma_{PWIA}$ is the cross-section calculated in the Plane Wave Impulse Approximation (PWIA) in which  all initial and final state  interactions  are neglected.   In the above expression for $T$, the cross-sections can be total cross-sections or differential cross-sections.

The  logic for color transparency to occur consists of three steps~\cite{ct,miller90,miller91}: a  high momentum transfer exclusive reaction proceeds by the formation of a small-sized or point-like configuration (PLC), a PLC has a small scattering amplitude because for a color neutral object, the sum of gluon emission amplitudes cancel, and a PLC expands as it moves. The second and third step requires the use of a  high-energy coherent process. The first step is the interesting assumption to be tested. 

 In reality, the cross-section $\sigma$ includes interactions of the incoming and outgoing particles.  These interactions  generally (but not always) lead  to a value for $\sigma$ which is smaller than $\sigma_{PWIA}$, and therefore $T<1$. Perturbative QCD predicts that in the limit of very large momentum transfer, $T\to1$, but the validity of 
 perturbative QCD is not a  necessary condition for color transparency to occur because confining interactions could lead to point-like-configurations~\cite{Frankfurt:1992dx,miller93}.
 
For the density of nuclear matter $\rho\simeq0.166\;fm^{-3}$, and proton-nucleon total cross-section $\sigma=40\;mb$ (for proton momentum greater than a few $GeV$), the mean-free path of the outgoing proton in the nucleus would be $l\simeq1.5\;fm$.  Thus for a nucleus of radius ~$3\;fm$ the outgoing proton would have a large probability of interacting with the other nucleons on its way out, and $T$ would be significantly smaller than $1$.   But the prediction of  color transparency is that the probability of the outgoing hadron interacting with the other nucleons on its way out is much smaller, and zero in the limit of very large momentum transfer.  In this case we would have $T\to1$.

At the present time, models must be used to account for the expansion effects.  The model used in this paper is called the ``quantum diffusion model"~\cite{liu88,dok91}.  In this model, the interaction cross-section of the outgoing object with the nucleons increases linearly with distance from the interaction point where the hard scatter occurred which produced the pointlike configuration (see \eq{sigeff}).  This model is derived from perturbative QCD~\cite{dok91}:  for a quark-antiquark system starting from a transverse size of zero, gluon exchange between the quark and antiquark proceeds until the system reaches the normal meson size.  It is shown in~\cite{dok91} that the transverse area of the system  (and hence its cross-section) increases linearly with distance traveled.  The ``naive" model of expansion would correspond to free quarks expanding from zero transverse size in both directions transverse to the momentum of the system.  In this case the transverse area of the system would increase as the square of the distance traveled~\cite{liu88}. The physics of the quantum diffusion model can also be captured by using a hadronic basis~~\cite{miller90,miller91,miller92,Miller:2012ce}.

The first experiment to search for effects of color transparency was in 1988 at Brookhaven National Laboratory~~\cite{carroll88}.  Quasi-elastic scattering of protons, $A(p,2p)A-1$, in various nuclei was observed, at incident proton momenta in the range of 6 to 12 GeV.  The transparency, as a function of the 4-momentum transfer squared $t$, was observed to increase as $\vert t\vert$ increased, up to a point, but then the transparency decreased after that as $\vert t\vert$ was increased further.  This behavior did not appear to agree with the predictions of color transparency, as the transparency should increase as $\vert t\vert$ is increased.  However, there may be other factors at work in the elementary $pp$ scattering cross-section, and several models were proposed to try to explain this behavior~~\cite{Miller:2007zzd}.  A later experiment ~~\cite{mardor98,leksanov01} obtained similar results

In the $(p,2p)$ reactions, in order for a small-sized configuration to be formed it is necessary to have 6 quarks all localized in a small region, which may have a very small probability.  The formation of a small-sized configuration may be more likely if fewer quarks are involved.  Thus quasi-elastic electron scattering ($e+A\to e+p+(A-1)$) may be a better candidate to observe color transparency.  Experiments have been performed at SLAC~~\cite{makins94,oneill95} and JLab~~\cite{garrow02} with a range of momentum-transfer squared $Q^2$ from $1$ to $8.1\;GeV^2$.  The results did not show any indication of color transparency.  The observations agreed with the standard calculation which assumes that the outgoing object is a normal-sized proton with the usual free-space value of its cross-section of interaction with the other nucleons.

There has been one experiment that can be said to show unambiguous evidence of color transparency.  This was the  nuclear diffractive dissociation of pions into dijets~{~\cite{Frankfurt:1993it,Frankfurt:1999tq}.  The result of the experiment~~\cite{aitala01} was a cross-section depending on $A$ as $A^{1.55}$~~\cite{Frankfurt:2000jm}, compared to $A^{2/3}$ which is what would be expected in the absence of CT effects.

Other candidate reactions are those involving electroproduction of pions~~\cite{Clasie:2007aa,co06,dutta03,miller2010}   or 
vector mesons~~\cite{Brodsky:1994kf,Gallmeister:2010wn,fassi12}.  Since the number of quarks involved in vector meson scattering or production is smaller than in proton scattering, the probability that all of the quarks involved are localized in a small space should be larger.   In contrast to the elastic reactions $p+A\to p+p+(A-1)$, $\pi+A\to\pi+A$, $\pi+A\to\pi+p+(A-1)$, etc., in electroproduction there are more parameters that may be varied, namely the virtual photon energy $\nu$ and virtuality $Q^2$.  These quantities, as well as a combination of them called the coherence length, $l_c=\frac{2\nu}{Q^2+m_V^2}$, can all affect the observed transparency.  The coherence length plays an especially important role, since by varying its value the transparency $T$ will vary even in the absence of any Color Transparency effects.  Thus to observe an actual CT effect, one must keep the coherence length fixed.            

There have been several searches for evidence of Color Transparency in electroproduction of $\rho$ mesons in nuclei.  At Fermilab in 1995~~\cite{adams95}, high energy muons were scattered from nuclei to produce $\rho$'s.  It was thought that CT was observed because the transparency, for a given $A$, increased as $Q^2$ was increased.  However, in this experiment the coherence length $l_c$ (see Sec. \ref{sec:vectormesonelectroproduction}) was not held constant as $Q^2$ was increased, so it is difficult to draw conclusions from their data.  A later experiment at DESY was conducted to explicitly measure the coherence length effect~~\cite{ackerstaff99}.  It was observed, as expected, that the transparency decreased as $l_c$ was increased, in $\rho$ electroproduction in $^{14}N$.  The $Q^2$ values for this experiment were such that no CT effects should occur, i.e. the produced object would interact with the full $\rho$-nucleon cross-section.  Hence any dependence of the transparency on $l_c$ was not an indication of CT.  This was a clear indication that any attempt to detect CT in vector meson electroproduction must look for effects while holding $l_c$ constant.  Another experiment at DESY~~\cite{airapetian03} was performed,  where the transparency as a function of $Q^2$ was measured for different values of $l_c$.  There appeared to be an increase in the transparency as $Q^2$ increased, although the number of events at each fixed value of $l_c$ was not large, and so better statistics are needed.  Finally, the most recent experiment to search for CT in $\rho$ production was at JLab~~\cite{fassi12}.  In this experiment, the coherence length varied from $\sim 0.5\;fm$ to $\sim 0.85\;fm$.  For this range of coherence length, the $q\bar{q}$ is produced essentially at the location of the nucleon that it scatters from, and so there are no Initial State Interactions.  The transparencies on $^{12}C$ and $^{56}Fe$ were measured for $Q^2$ from $1.0$ to $2.3\;GeV^2$.  The transparencies appeared to show an increase with $Q^2$, although the kinematic range covered was small. 

In this paper we calculate the transparency $T$ for the proton knockout reaction $\gamma^*+A\to \rho +p+(A-1)$, in both the standard Glauber model~~\cite{glaub59,glaub67} (which does not account for Color Transparency effects) and in the Glauber model modified to include CT effects. 
The aim is to see if insight can be gained by detecting the outgoing proton.

  The paper is organized as follows.  Sec. \ref{sec:vectormesonelectroproduction} briefly reviews the electroproduction of a vector meson from a single nucleon, including the coherence length and formation time.  Sec. \ref{sec:second} discusses the Glauber formalism for particle production which will be used in the calculations.  In this section we calculate the production amplitude for the case where the residual nucleus is a one-hole state of the initial nucleus, in the shell model, for both the case of neglecting CT effects and the case of including CT effects.  The result is expressed in terms of the missing momentum $\mathbf{p}_m$ of the reaction.   In Sec. \ref{sec:CT} we discuss the modification to the Glauber result which we make in order to include the effects of Color Transparency.
In Sec. \ref{sec:transparency} the transparency $T(\mathbf{p}_m)$ is calculated using shell-model wavefunctions, for $\mathbf{p}_m=0$.  Results are presented for $A=12$ and $A=40$, for a range of values $Q^2$ and $l_c$.  In Sec. \ref{sec:third} the ``integrated transparency" is calculated.  This is the transparency where the numerator and denominator of \eq{transpardef} are integrated over a domain of $\mathbf{p}_m$.  This is the quantity of more direct experimental significance, rather than $T(\mathbf{p}_m)$ for a particular value of $\mathbf{p}_m$.  The integrated transparency was calculated for $A=12$ and $A=40$, for both fixed $t$  and varying $Q^2$ (where $t$ is the 4-momentum-transfer squared to the proton), and also fixed $Q^2$ and varying $t$.  It is shown that the effects of CT can be seen even for small values of $Q^2$, if $t$ is large enough.  Sec. \ref{sec:conclusion} summarizes.

 In~~\cite{miller08} the cross-section $\frac{d\sigma}{dt}$ for semi-inclusive electroproduction of $\rho$ mesons was calculated, including effects of color transparency, and using the Glauber model.  In that paper the decay of the $\rho$ to pions was accounted for, using the Glauber model.  It was shown that the effects of $\rho$ decay are small for large photon energies; hence in this paper we neglect them.

\section{Electroproduction of a vector meson on a single nucleon}
\label{sec:vectormesonelectroproduction}

High-energy electroproduction of vector mesons from a nucleon can be described in terms of quark degrees of freedom (QCD) or hadronic degrees of freedom (e.g. Vector Meson Dominance~~\cite{feyn72,yennie78}).  In the two descriptions, the incident virtual photon first fluctuates into either a virtual quark-antiquark pair or into a virtual vector meson, respectively.  The virtual $q\bar{q}$ pair or vector meson then scatters elastically from the nucleon.  The momentum transfer involved puts the  virtual $q\bar{q}$ pair or vector meson on the mass-shell of the final observed vector meson.  The virtual $q\bar{q}$ pair or vector meson then evolves over time (since it's not an eigenstate of the strong force Hamiltonian) to form the final observed vector meson. 
In the quark picture, the transverse size $r_{\perp}$ of the $q\bar{q}$ that is produced by the virtual photon goes as $r_{\perp}\simeq 1/Q$~~\cite{Miller:2007zzd}, so the larger $Q$ is, the smaller is the size of the produced $q\bar{q}$.  In the limit of $Q\to\infty$ the size goes to zero:  a point-like configuration.  Thus for large $Q^2$ the produced object should have vanishing interactions with the other nucleons and the transparency should approach $1$.

In both descriptions there are two time-scales (or length scales, since the velocity of the vector meson is approximately $c$) which are of relevance (Fig. \ref{fig:coherlength}).  The first is the ``coherence length", $l_c$, which is the distance that the virtual hadronic fluctuation of the photon can travel in the LAB frame (target nucleon or nucleus at rest)~\cite{Miller:2007zzd}.  The energy-time uncertainty relation can be used to determine this distance.  For a photon of energy $\nu$ and 4-momentum-squared $-Q^2$, it is given by
\be
l_c=\frac{2\nu}{Q^2+m_V^2}.
\ee
where $m_V$ is the mass of the vector meson.  The other time scale of relevance is called the ``formation time".  The formation time is the time scale over which the virtual meson or $q\bar{q}$ pair develops into the final real vector meson state, after scattering from the nucleon.  The scattering with the nucleon puts the virtual meson or $q\bar{q}$ pair onto the mass shell of the vector meson.  At the time of scattering the transverse size of the $q\bar{q}$ is small, and as it propagates away it evolves into the final meson state.  This time can be estimated by considering the on-mass-shell small-size $q\bar{q}$ pair as a superposition of hadron states, namely the final real vector meson state and the next higher-mass meson state~~\cite{Miller:2007zzd}.  Then the energy-time uncertainty principle in the rest frame of the outgoing meson gives
$\Delta t=\frac{1}{m_{V'}-m_V},
$
while in the LAB frame this is time-dilated  so the formation time or length $l_h$ in the LAB (assuming $\beta\simeq c$) is
\be
l_h=\frac{2\;p_V}{m_{V'}^2-m_V^2}
\ee
where $p_V$ is the outgoing vector meson's momentum.

For vector meson production in a nucleus, while the virtual hadron or $q\bar{q}$ is propagating over the distance $l_c$ it may interact with nucleons and be absorbed, before it has a chance to undergo the interaction which puts it on mass-shell.  These Initial State Interactions (ISI) therefore affect the measured production cross-section in the nucleus.  In general, as $l_c$ increases, the probability of absorption increases and so the measured production cross-section in a given nucleus should decrease.  Thus the production cross-section at low energy (small $\nu$) should be larger than the production cross-section at high energy (large $\nu$), for a given $Q^2$.  Or conversely, for a given $\nu$, as $Q^2$ is increased, $l_c$ will decrease and therefore the measured production cross-section should increase.  This effect mimics the effect of Color Transparency.  Therefore in order to detect effects of CT, the coherence length should be kept fixed in a given experiment.

\begin{figure}[htb]
     \begin{center}

   \includegraphics[width=4.8in,height=2.5in]{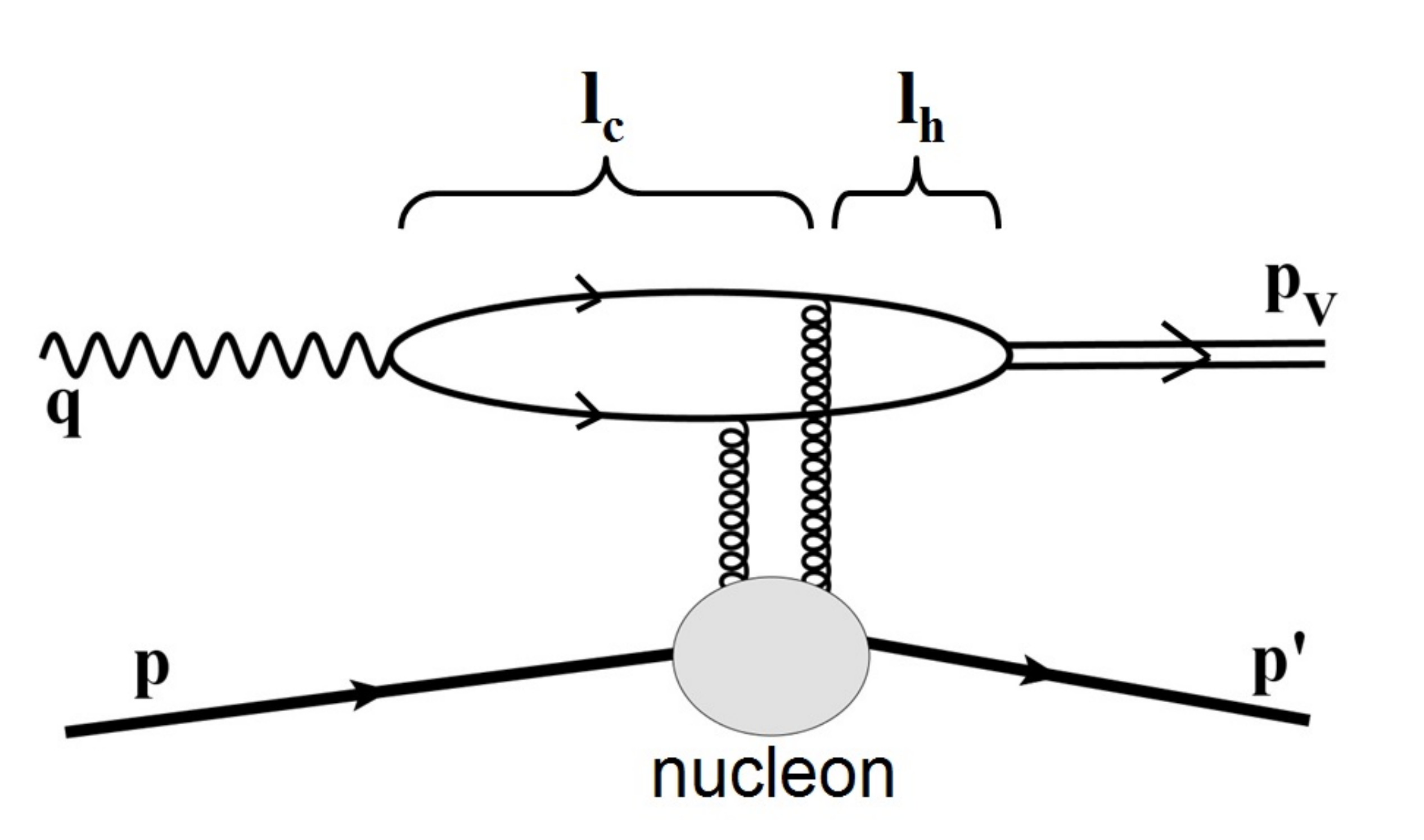}
    \end{center}
    \caption{%
        Coherence length ($l_c$) and formation length ($l_h$) for vector meson electroproduction.  The incoming photon dissociates into a $q\bar{q}$ pair which then interact with the nucleon by exchanging gluons.  
     }%
   \label{fig:coherlength}
\end{figure}

\newcommand{\bfr}{{\bf r}}

\newcommand{\ben}{\begin{displaymath}}
\newcommand{\een}{\end{displaymath}}

\newcommand{\bfp}{{\bf p}}

\section{Production amplitude ignoring effects of CT}
\label{sec:second}

We consider the $\rho$ meson production process \bea\gamma^*+A\to \rho+p+(A-1)^*,\label{reac}\eea where $(A-1)^*$ means the final $(A-1)$-nucleon system is allowed to be in any final state.  In this section we   calculate the amplitude for this process ignoring any effects of Color Transparency.  
Since we are interested in large incident photon energy, the Glauber model of high-energy hadron-nucleus scattering~~\cite{glaub59}  is applicable.  The Glauber model is a multiple-scattering model which is valid under certain conditions:  1) the energy of the incident particle must be very large, compared to the binding energy of the nucleons in the target nucleus; 2)  the scattering angle of the projectile is small.  Under these conditions the momentum transfer from the projectile is mostly transverse, and so the longitudinal momentum transfer is neglected; the energy transfer from the projectile is also small, and so the energy transfer is neglected also.   In the Glauber model, the nucleon positions are fixed in place during the time that the projectile traverses the nucleus (the ``frozen" approximation).  The projectile is assumed to scatter at most once from any individual nucleon.  In between scattering events the projectile travels in a straight line.  The Glauber result for the scattering amplitude is a sum of terms representing the possible multiple-scatterings of the projectile.   In the case of particle production, the Glauber model is modified~~\cite{marg68,kopel96} to take into account the longitudinal momentum transfer in the production process on a nucleon, which is necessarily non-zero due to the difference in mass of the incident particle (here the $\gamma^*$) and the produced particle (the $\rho$).  The Glauber model does not take into account the Fermi motion of the nucleons; for a projectile of high energy the Fermi motion should matter little.  In order to calculate the scattering cross-section in the Glauber model, only knowledge of the free-space hadron-nucleon scattering amplitudes and the wavefunctions of the target system is required.

\subsection{Glauber formalism for particle production}

For the reaction $\gamma^*+A\to \rho +A^*$, where $A^*$ represents any final state of the $A$-nucleon system, we define $\mathbf{k}$ as the incident virtual photon 3-momentum, $\mathbf{k}^{\prime}$ is the 3-momentum of the outgoing $\rho$, and $\mathbf{q}\equiv\mathbf{k}-\mathbf{k}^{\prime}$ is the 3-momentum transfer.  The coordinate system is such that $\mathbf{k}$ defines  the positive $z$-direction.  
In the Glauber theory, the scattering amplitude, for a transition from the initial target ($A$-nucleon) internal state $\vert i\rangle$ to the final $A$-nucleon internal state $\vert f\rangle$, is given by
\be
\label{glauberamplitude}
F_{fi}(\mathbf{q})=\frac{ik}{2\pi} \int d^2b\;e^{i\mathbf{q}\cdot \mathbf{b}}\;\langle f\vert \Gamma_{tot}(\mathbf{b},\mathbf{r}_1,...,\mathbf{r}_A)\vert i\rangle\;
\ee
where {\bf b} is the impact parameter and    profile operator $\Gamma_{tot}$ given by~~\cite{marg68,kopel96}
\be
\label{Vgammatot}
\Gamma_{tot}(\mathbf{b},\mathbf{r}_1,\ldots,\mathbf{r}_A)=\sum_{j=1}^{A} \Gamma^{\gamma}(\mathbf{b}-\mathbf{s}_j)e^{iq_Lz_j} \prod_{m\neq j}^A\bigl [1-\Gamma(\mathbf{b}-\mathbf{s}_m)\theta(z_m-z_j)\bigr] . 
\ee
The coordinate $\mathbf{r}_j$ is the position vector of the $j$th nucleon:  $\mathbf{r}_j=(\mathbf{s}_j,z_j)$. The   vector $\mathbf{s}_j$ is the projection of $\mathbf{r}_j$ in the plane transverse to the $z$-axis.  The set of nucleon coordinates $\lbrace \mathbf{r}_j\rbrace$ are the internal coordinates of the $A$-nucleon system, and hence are relative coordinates.  For the $A$-nucleon system there are $A-1$ independent coordinates.  The terms ``transverse" and ``longitudinal" are in relation to the $z$-axis:  ``transverse" means in the plane transverse to the $z$-axis, while ``longitudinal" means parallel to the $z$-axis.

In \eq{Vgammatot}, $q_L$ is the longitudinal momentum transfer to the nucleon on which the forward production of the vector meson occurs.   For the case of $\gamma^*+N\to \rho +N$ at high energy, and forward production of the rho meson
\be
q_L=\frac{Q^2+M^2}{2\nu}.
\ee

The 2-body profile function $\Gamma^{\gamma}$ is related to the vector meson production amplitude from a single nucleon, $f^{\gamma V}(\mathbf{q})$ (i.e. for the process $\gamma^*+N\to V+N$), where $\mathbf{q}$ is the transverse momentum transfer, by 
\be
\label{ampgammaV}
f^{\gamma V}(\mathbf{q})=\frac{ik}{2\pi}\int d^2b\; e^{i\mathbf{q}\cdot\mathbf{b}}\;\Gamma^{\gamma}(\mathbf{b})
\ee 
Thus we have
\be
\Gamma^{\gamma}(\mathbf{b})=\frac{1}{2\pi ik}\int d^2q\; e^{-i\mathbf{q}\cdot\mathbf{b}}f^{\gamma V}(\mathbf{q}),
\ee 
which gives $\Gamma^{\gamma}$ in terms of $f^{\gamma V}$.   The 2-body profile function $\Gamma(\mathbf{b})$ is related to the scattering amplitude for elastic vector meson-nucleon scattering, $f(\mathbf{q})$, by  
\be
\label{ampVN}
f(\mathbf{q})=\frac{ik}{2\pi}\int d^2b \;e^{i\mathbf{q}\cdot\mathbf{b}}\;\Gamma(\mathbf{b}),
\ee 
and also
\be
\label{gammVN}
\Gamma(\mathbf{b})=\frac{1}{2\pi ik}\int d^2q\; e^{-i\mathbf{q}\cdot\mathbf{b}}f(\mathbf{q}).
\ee        
The total profile operator $\Gamma_{tot}$, \eq{Vgammatot}, thus represents production of the vector meson on a nucleon at $(\mathbf{s}_j,z_j)$, with longitudinal momentum transfer $q_L$, followed by elastic re-scatterings of the produced meson on the other nucleons (up to a maximum of $A-1$ rescatterings).  The factor $\theta(z_m-z_j)$ ensures that any elastic scattering of the produced vector meson  on a nucleon at longitudinal position $z_m$      occurs \underline{after} the meson has been produced on the nucleon at position $z_j$ (since the produced vector meson's velocity is mostly in the positive $z$-direction).  The total amplitude for production of the vector meson from the nucleus, including the effects of rescattering of the produced vector meson from individual nucleons, is the 2-dimensional Fourier transform of the matrix element of the operator $\Gamma_{tot}$ between the initial and final internal states of the $A$-nucleon system, \eq{glauberamplitude}, and is a sum of terms representing no elastic rescattering of the vector meson, one elastic rescattering, two rescatterings, etc., up to a maximum of $A-1$ rescatterings.

\subsection{Production amplitude}
\label{sec:amplitude}

We   calculate the differential cross-section for the process in which a single nucleon is knocked out of the nucleus. The  final states of the residual nucleus (the ($A-1$)-nucleon system)   are one-hole states of the initial nucleus. 
We use shell-model wavefunctions for the initial target state, in which the single-particle wavefunction  of the knocked out  nucleon is denoted by
 $\phi_n(\bfr_1)$.
The final $A$-nucleon state   is taken to be one in which the nucleon of wave function $\phi_n(\bfr_1)$ is replaced by the scattering wave function
$\chi_{\mathbf{p}}(\mathbf{r}_1)$  for the proton of momentum $\mathbf{p}. $ 
The subscript $n$ defines a state in which a proton in the single-particle state $n$ is removed from the initial ground state wave function.
 We shall be concerned with states in which the  energy and momentum transfer to  the outgoing proton is high enough so that the  eikonal wave function can be used:
\be
\chi_{\mathbf{p}}(\mathbf{r}_1)=e^{i\mathbf{p}\cdot\mathbf{r}_1}e^{-\frac{1}{2}\int_0^{\infty}ds\; \sigma_{tot}^{pN}\rho(\mathbf{r}_1+s\hat{\mathbf{p}})}\equiv e^{i\mathbf{p}\cdot\mathbf{r}_1} e^{-\frac{1}{2}\alpha_{\mathbf{p}}(\mathbf{r}_1)} .
\ee
Here $\sigma_{tot}^{pN}$ is the total cross-section for proton-nucleon scattering, and $\rho(\mathbf{r})$ is the nucleon number density for the residual nucleus.
$\chi_{\mathbf{p}}(\mathbf{r}_1)$ represents scattering of the outgoing proton in the optical potential due to the other $A-1$ nucleons, which are in the bound state $\Psi_{A-1}^f$.  Therefore $\rho$ in the exponential should in principle depend on the final state $f$ of the residual nucleus.  We will assume, however, that $\rho$ is the same as the nucleon density of the initial nucleus, which should be approximately correct for final states which are one-hole states or small excitations thereof, for large $A$.

We compute  the scattering amplitude for the stated  final state by 
using the discussed initial and final states in 
  Eq.~\ref{glauberamplitude}.  We assume that the final state proton is created in the 
production of the $\rho$ or is created by a $\rho$-proton final state interaction. In that case, we find
\be
\label{Ffi51}
F_{fi}^{(n)}=\frac{ik}{2\pi}\int d^2b e^{i\mathbf{q}\cdot\mathbf{b}}\int d^3r_1  \chi_p^*(\mathbf{r}_1)\phi_n(\mathbf{r}_1)
\left(\Gamma^{\gamma}_{b1}e^{iq_Lz_1}g_1(\mathbf{b})
-\Gamma_{b1}\int d^3r_2\; \rho(\mathbf{r}_2)\; \Gamma^{\gamma}_{b2}\;e^{iq_Lz_2}\;\theta_{12}\; g_2(\mathbf{b})\right),
\ee
in which 
 we use the definitions  $\Gamma_{bj}^{\gamma}\equiv\Gamma^{\gamma}(\mathbf{b}-\mathbf{s}_j)$, $\Gamma_{bk}\equiv\Gamma(\mathbf{b}-\mathbf{s}_k)$,  $\theta_{kj}\equiv\theta(z_k-z_j)$, and 
\be
g_1(\mathbf{b})\equiv \Bigl[1-\int d^2s \int_{z_1}^{\infty}dz \rho_1(\mathbf{s},z)\Gamma(\mathbf{b}-\mathbf{s})\Bigr]^{A-1},
\ee
\be
g_2(\mathbf{b})\equiv \Bigl[1-\int d^2s \int_{z_2}^{\infty}dz \rho_1(\mathbf{s},z)\Gamma(\mathbf{b}-\mathbf{s})\Bigr]^{A-2},
\ee
where $\rho_1$ is related to the nucleon number density $\rho$ by $\rho(\mathbf{r})=(A-1)\rho_1(\mathbf{r})$, with normalization $\int d^3r \rho_1(\mathbf{r})=1$.

To obtain a tractable form for the amplitude $F_{fi}^{(n)}$, we can utilize the fact that the profile functions are sharply peaked as their arguments vary while the other quantities appearing in the expression for  $F_{fi}^{(n)}$ (i.e.  $\rho$, $g_1$, and $g_2$), are relatively slowly varying. This is because the range of the profile function is of the order of the size of the nucleon, while the other functions vary over the size of the nucleus.   For a slowly varying function $f(\mathbf{r})$ we thus have, to good approximation, 
\be
\label{approxGammaint}
\int d^2s f(\mathbf{s},z)\Gamma(\mathbf{s}-\mathbf{a})\simeq f(\mathbf{a},z)\int d^2s \Gamma(\mathbf{s}-\mathbf{a})
\ee
and similarly for $\Gamma^{\gamma}(\mathbf{s}-\mathbf{a})$.

Using this approximation in $g_1(\mathbf{b})$ and $g_2(\mathbf{b})$, we obtain in the large-$A$ limit,
\be
g_1(\mathbf{b})\simeq e^{-\frac{1}{2}\sigma_{tot}^{VN}\;T_1(\mathbf{b})},
\ee
\be
g_2(\mathbf{b})\simeq e^{-\frac{1}{2}\sigma_{tot}^{VN}\;T_2(\mathbf{b})}
\ee
where $T_j(\mathbf{b})\equiv \int_{z_j}^{\infty} dz' \rho(\mathbf{b},z')$ is called the ``partial thickness function".  Note that the optical theorem was used to relate the forward elastic scattering amplitude $f(0)=\frac{ik}{2\pi}\int d^2s\;\Gamma(\mathbf{s})$ to the total vector meson-nucleon cross-section $\sigma_{tot}^{VN}$.

Using the above approximations we find 
 the result:
\begin{equation}
\label{Vresult}
\begin{split}F_{fi}^{(n)}=& \int d^2s_1  d z_1  e^{-i\mathbf{p}_m\cdot\mathbf{r}_1}   e^{-\frac{1}{2}\alpha_p(\mathbf{r}_1)}  \phi_n(\mathbf{r}_1)\\
& \times\Bigl(f^{\gamma V}(\mathbf{q}) e^{-\frac{1}{2}\sigma_{tot}^{V N}\;T_1(\mathbf{s}_1)} - \frac{2\pi}{ik}f^{\gamma V}(0)  \int_{-\infty}^{z_1} dz_2\; \rho(\mathbf{s}_1 ,z_2) \;e^{iq_L(z_2-z_1)}\;  e^{-\frac{1}{2}\sigma_{tot}^{V N}\;T_2(\mathbf{s}_1)} f(\mathbf{q})\Bigr).\end{split}
\ee
The result for $F_{fi}^{(n)}$ depends on  the missing momentum $\mathbf{p}_m$:
\be
\mathbf{p}_m\equiv \mathbf{p} -\mathbf{k}+\mathbf{k}'=\mathbf{p}_{\perp}-\mathbf{q} + (p_z-q_L) \hat{\mathbf{z}}\label{pmdef}
\ee
where $\mathbf{p}$ is the momentum of the outgoing proton. 
\begin{figure}[tbp]
     \begin{center}
        \subfigure[First term of \eq{Vresult} ]{%
            \label{fig:pictorialrep1}
        }%
          \includegraphics[width=0.5\textwidth,height=1.8in]{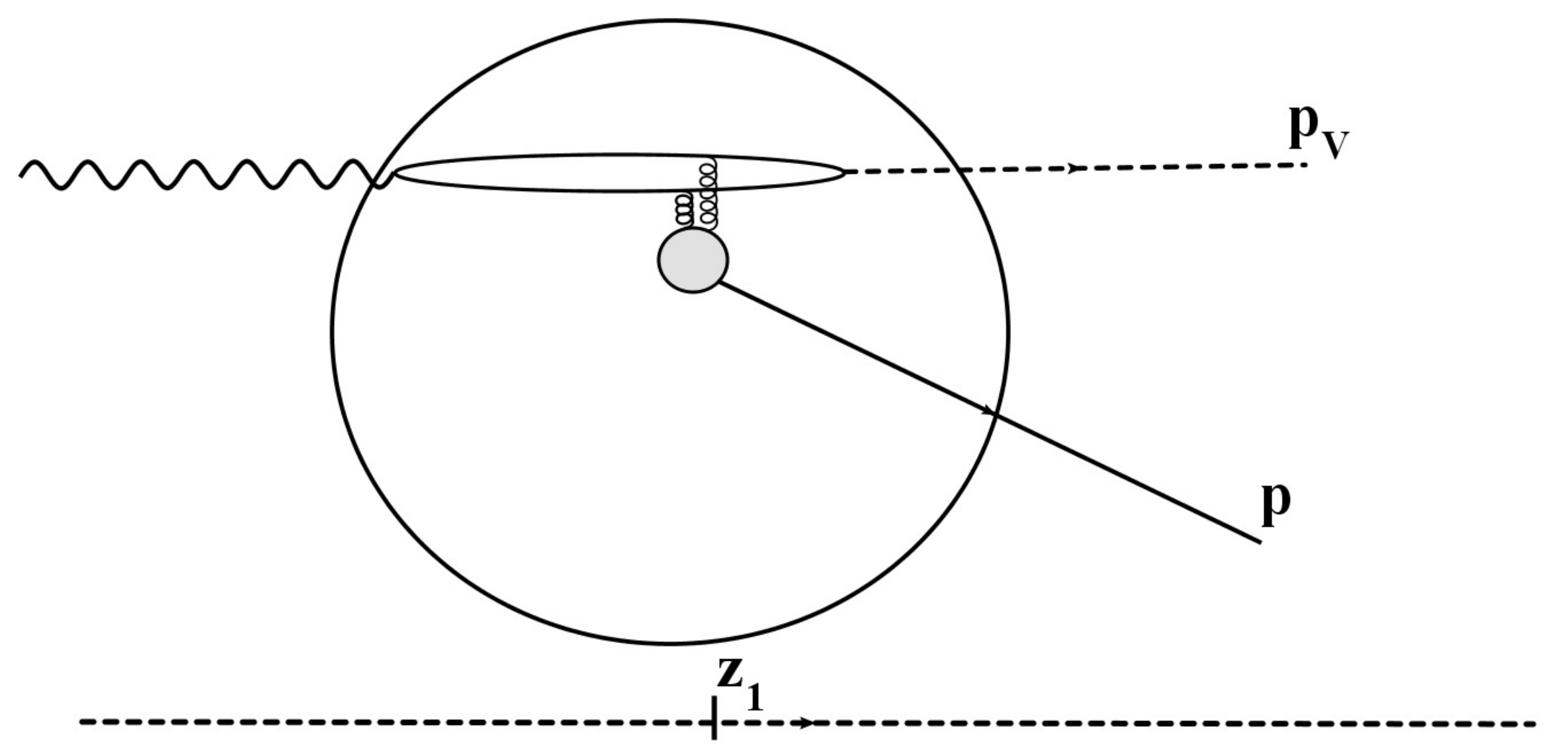}
        \subfigure[Second term of \eq{Vresult}]{%
           \label{fig:pictorialrep2}
           \includegraphics[width=0.5\textwidth,height=1.8in]{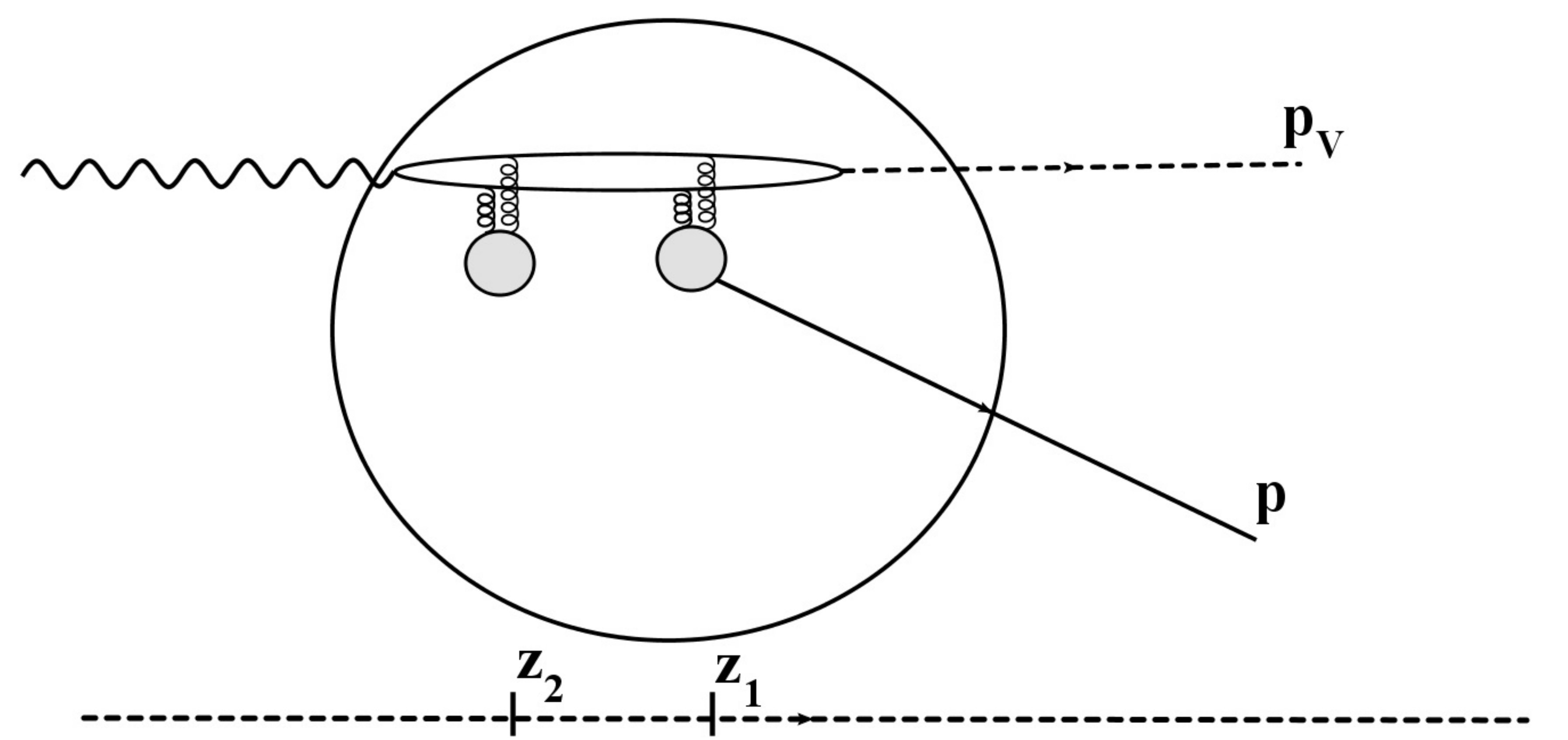}
        }\\ 

    \end{center}
    \caption{%
        Pictorial representation of the two terms in the amplitude of \eq{Vresult}.  
     }%
   \label{fig:pictorialrep}
\end{figure}

The physical interpretation of the two terms in \eq{Vresult} is as follows.  The first term in parentheses corresponds to production of the vector meson on nucleon 1 at position $(\mathbf{s}_1,z_1)$ with transverse momentum transfer $\mathbf{q}$, nucleon 1 being therefore knocked out. The second term in parentheses corresponds to forward production of the vector meson on nucleon 2 at position $(\mathbf{s}_1,z_2)$; the produced meson then propagates in the $z$-direction until the point $(\mathbf{s}_1,z_1)$  where it scatters elastically from nucleon 1 with transverse momentum transfer $\mathbf{q}$ to nucleon 1, nucleon 1 being knocked out.  In both cases the vector meson suffers attenuation beginning at the point where it is created as a physical meson through interaction with a nucleon (either at $(\mathbf{s}_1,z_1)$ for the first term or at $(\mathbf{s}_1,z_2)$ for the second term), while the proton suffers attenuation beginning at the point $\mathbf{r}_1=(\mathbf{s}_1,z_1)$ where it was located when the vector meson struck it.  The total amplitude is the sum of these two amplitudes; hence the square of the amplitude contains interference between the two amplitudes.

The result   \eq{Vresult} is the Glauber theory result for the scattering amplitude for $\gamma^*+A\to\rho+p+(A-1)^*$, for the case where the final state of the residual nucleus is a one-hole state of the initial nucleus.  To obtain the differential cross-section, summed over all one-hole states, we would square $F_{fi}^{(n)}$, multiply by the appropriate phase-space and flux factors, and then sum over $n=1$ to $A$.  For the high-energy case we are considering, we may consider the energies of the outgoing particles to be essentially independent of $n$.  In that case, the phase-space and flux factors are independent of $n$, and so we may just sum $\vert F_{fi}^{(n)}\vert^2$ over $n$.   

The result for $\sum_n \vert F_{fi}^{(n)}\vert^2$, where $n$ is summed only over one-hole states, is identical to the result one would obtain if instead one summed over \underline{all} final states of the residual nucleus (the incoherent cross-section) but only kept the terms corresponding to a single rescattering of the produced vector meson on a proton, and neglected terms where the vector meson rescatters two or more times on different nucleons.  The experimental situation, wherein the recoiling nucleus is not detected, corresponds to summing over all final states of the residual nucleus.  However, because of the exclusive nature of the reaction, if $\mathbf{p}_m$ is small, then the outgoing proton's momentum $\mathbf{p}\simeq\mathbf{q}$ and so only a single rescattering of the $\rho$ can have occurred, where the entire momentum transfer $\mathbf{q}$ was delivered to the detected proton.  Multiple rescattering terms in this case should be negligible, and so we need only sum $ \vert F_{fi}^{(n)}\vert^2$ over one-hole final states.  This implies that the transparency $T$ using the result \eq{Vresult}  will show very little dependence on the 4-momentum-transfer-squared $t\simeq -\mathbf{q}^2$.  

\section{Inclusion of Color Transparency effects}
\label{sec:CT}

Effects of Color Transparency can be incorporated into the result \eq{Vresult} by including position dependent cross-sections from the quantum diffusion model~~\cite{liu88,dok91}.  In this model, the total cross-section of interaction of the outgoing hadrons with a nucleon in the nucleus is~~\cite{liu88}
\be
\label{sigeff}
\sigma^{eff}_{hN}(z,t)=\sigma^{tot}_{hN}\Biggl[ \theta(l_h-z)\; \Bigl[\frac{z}{l_h}+\frac{n^2\langle k_t^2\rangle}{\vert t\vert}\Bigl (1-\frac{z}{l_h}\Bigr) \Bigr]  +\theta(z-l_h)\Biggr].  
\ee
Here $z$ is the distance the hadron has traveled from the point where the hard hadron-nucleon interaction (with 4-momentum-transfer-squared $t$) occurred (Fig. \ref{fig:hardscatter}), 
$\sigma^{tot}_{hN}$ is the free-space total hadron-nucleon cross-section,  $n$ is the number of valence quarks of the hadron, and $\langle k_t^2\rangle^{1/2}$ is the average transverse momentum of the quark in the hadron (taken to be  $\langle k_t^2\rangle^{1/2}=0.35$ GeV).  Thus $\frac{n^2\langle k_t^2\rangle}{\vert t\vert}\sigma^{tot}_{hN}$ is a measure of the transverse size of the hadron at the time of collision.  The parameter $l_h$ (the formation length) is the distance the hadron travels after the collision until it reaches its normal size.  This is estimated as $l_h\simeq\frac{1}{E_n-E_h}\simeq\frac{2 p_h}{M_n^2-M_h^2}$, where $M_n$ is the mass of a typical intermediate state $n$ of the hadron~~\cite{liu88}.  In principle the quantity $l_h$ can be different for the meson  and the proton, but since the relation $l_h\simeq\frac{1}{E_n-E_h}\simeq\frac{2 p_h}{M_n^2-M_h^2}$ is only an estimate, we take here $M_n^2-M_N^2=M_n^2-M_{\rho}^2=0.7\;GeV^2$ for both $l_{\rho}$ and $l_p$~~\cite{miller06}.      

\begin{figure}[htb]
     \begin{center}

            \includegraphics[width=4.8in,height=2.5in]{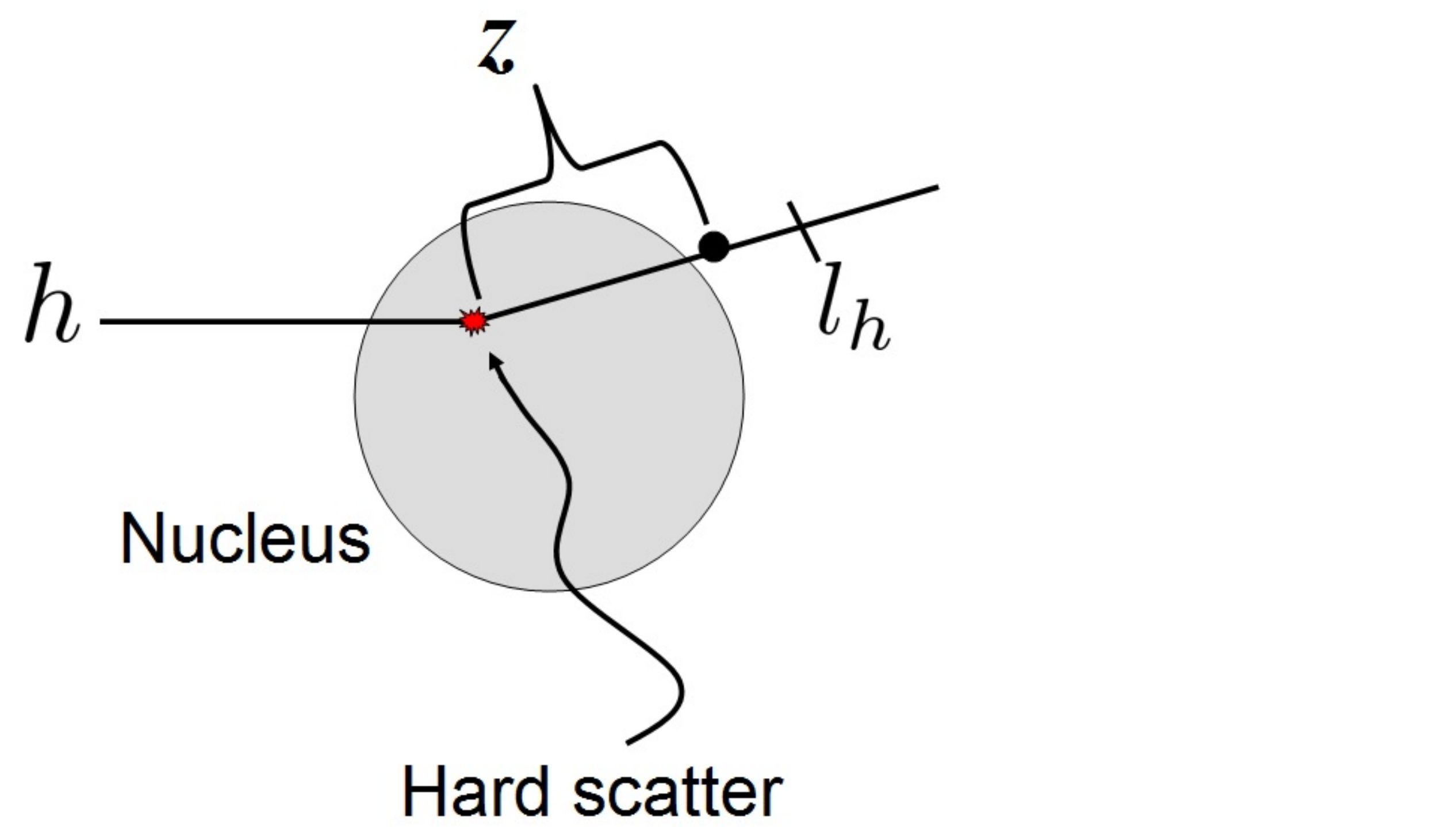}

    \end{center}
    \caption{%
        Formation length ($l_h$) for vector meson production. $z$ is the distance of the outgoing hadron from the point where the hard scattering occurred.
     }%
   \label{fig:hardscatter}
\end{figure}

The expression \eq{sigeff} is used for the cross-sections that appear in the exponentials in \eq{Vresult}.  The amplitudes $f^{\gamma V}(\mathbf{q})$ and $f^{\gamma V}(0)$ that appear in \eq{Vresult} are the same as the measured free-space production amplitudes.  However, the elastic rescattering amplitude $f(\mathbf{q})$ must be modified to include the effects of Color Transparency.  For large enough $Q^2$, the $q\bar{q}$ pair produced at the point $(\mathbf{s}_1,z_2)$ will be in a pointlike configuration; it will then expand as it propagates, and scatter elastically from a nucleon at $z_2$; if $z_2$ is close enough to $z_1$, the scattering amplitude $f(\mathbf{q})$ of the $q\bar{q}$ pair on the nucleon will be smaller than that of a normal $\rho$ meson.   Therefore the scattering amplitude $f(\mathbf{q})$ in \eq{Vresult} should be replaced by~~\cite{Frankfurt:1994kk}
\be
\label{eq:fplc}
f^{PLC}(z_1-z_2,\mathbf{q},Q^2)=f(\mathbf{q})\frac{\sigma^{eff}_{VN}(z_1-z_2,Q^2)}{\sigma^{tot}_{VN}}\frac{G_V\Bigl(t\;\frac{\sigma^{eff}_{VN}(z_1-z_2,Q^2)}{\sigma^{tot}_{VN}}\Bigr)}{G_V(t)}
\ee
where $G_V(t)$ is the $\rho$-meson form factor, and $t\simeq -\mathbf{q}^2$, and $f(\mathbf{q})$ is the measured free-space elastic $\rho$-nucleon scattering amplitude.  This form for $f^{PLC}$ is derived using the optical theorem (and assuming $f(0)$ is pure imaginary) together with the empirical result~~\cite{collins84,Frankfurt:1994kk} that the differential cross-section for hadron-nucleon scattering satisfies 
\be
\frac{d\sigma^{hN\to hN}}{dt}\sim G_h^2(t)G_N^2(t).
\ee 
in terms of the form factors of the $h$ and $N$.

Thus the result for the scattering amplitude including Color Transparency effects is
\begin{equation}
\label{VresultCT}
\begin{split}&F_{fi}^{(n)}=\int d^2s_1  d z_1  e^{-i\mathbf{p}_m\cdot\mathbf{r}_1}   e^{-\frac{1}{2}\alpha_{\mathbf{p}}(\mathbf{r}_1)}  \phi_n(\mathbf{r}_1)\\
& \times\Bigl(f^{\gamma V}(\mathbf{q}) e^{-\frac{1}{2}\alpha_V(\mathbf{s}_1,z_1)} - \frac{2\pi}{ik}f^{\gamma V}(0)  \int_{-\infty}^{z_1} dz_2\; \rho(\mathbf{s}_1 ,z_2)\; e^{iq_L(z_2-z_1)}\;  e^{-\frac{1}{2}\alpha_V(\mathbf{s}_1,z_2)} f^{PLC}(z_1,z_2,\mathbf{q},Q^2)\Bigr)\end{split}.
\ee
where
\be
\alpha_{\mathbf{p}}(\mathbf{r}_1)= \int_0^{\infty}\sigma^{eff}_{pN}(s,t)\rho(\mathbf{r}_1+s\;\hat{\mathbf{p}})ds
\ee
\be
\alpha_V(\mathbf{s}_1,z_1)=\int_{z_1}^{\infty} dz^{\prime}\sigma^{eff}_{V N}(z^{\prime}-z_1,t)\rho(\mathbf{s}_1,z^{\prime})
\ee
\be
\alpha_V(\mathbf{s}_1,z_2)=\int_{z_2}^{\infty} dz^{\prime}\sigma^{eff}_{V N}(z^{\prime}-z_2,Q^2)\rho(\mathbf{s}_1,z^{\prime}).
\ee
These expressions for $\alpha_V$ reflect the fact that the transverse size of the initial $q\bar{q}$ (at $z_2$) is determined by $1/Q^2$, while the transverse size of the outgoing  $q\bar{q}$ and proton, after the hard scatter from the proton at $(\mathbf{s}_1,z_1)$, is determined by $1/\vert t\vert$.

\section{Transparency}
\label{sec:transparency}

For the proton knockout reaction, the transparency $T$ is defined as the ratio of the measured 5-fold differential cross-section to the differential cross-section calculated in the Plane Wave Impulse Approximation (PWIA)~~\cite{garrow02, oneill95, makins94,benhar96}.  This can be evaluated at a specific kinematic point, i.e. a particular value of the missing momentum $\mathbf{p}_m$, or it can be the ratio of the integrated cross-sections, integrated over some domain $\cal{D}$ of $\mathbf{p}_m$.  Thus
\be
\label{transpdef}
T(\mathbf{p}_m)=\frac{\frac{d\sigma}{dE'd\Omega'  d\Omega_p}}{\frac{d\sigma_{PWIA}}{dE'd\Omega'  d\Omega_p}},
\ee 
or
\be
\label{integratedT}
T_{\cal{D}}=\frac{\int_{\cal{D}}d^3p_m\frac{d\sigma}{dE'd\Omega'  d\Omega_p}}{\int_{\cal{D}}d^3p_m \frac{d\sigma_{PWIA}}{dE'd\Omega'  d\Omega_p}}.
\ee 
We will call the latter the ``integrated transparency".
At a given value of $\mathbf{p}_m$, the kinematic factors in the cross-sections cancel in the ratio \eq{transpdef}.  We are summing the cross-sections over the one-hole final states, labeled by $n$, from $n=1...A$.  If we neglect the dependence of $F_{fi}^{(n)}$ on the binding energy of the state $n$ (which is valid for large momentum $\mathbf{p}$ of the outgoing proton) then the sum of the cross-sections is proportional to $\sum_{n=1}^A \vert F_{fi}^{(n)}\vert^2$, and so we have
\be
\label{transp}
T(\mathbf{p}_m)=\frac{\sum_{n=1}^A \vert F_{fi}^{(n)}\vert^2}{\sum_{n=1}^A \vert F_{fi}^{(n)}\vert_{PWIA}^2}.
\ee
The PWIA value of $ F_{fi}^{(n)}$ is obtained from \eq{Vresult} or \eq{VresultCT} by setting the exponential attenuation factors equal to 1 and setting the second term in parentheses equal to zero (which means that all rescattering of the produced vector meson is neglected).  Using the expression \eq{Vresult} for $ F_{fi}^{(n)}$ in \eq{transp} gives the Glauber theory prediction for the nuclear transparency $T$ (called here the ``Glauber result"), while using \eq{VresultCT} for $ F_{fi}^{(n)}$ in \eq{transp} gives the prediction for the nuclear transparency $T$ including effects of Color Transparency (called here the ``CT result").  In both cases the denominator of \eq{transp} is just 
\be
\label{denom}
\sum_{n=1}^A \vert F_{fi}^{(n)}\vert_{PWIA}^2=\vert f^{\gamma V}(\mathbf{q})\vert^2 \;\sum_{n=1}^A \Bigl\vert\int d^2s_1  d z_1  e^{-i\mathbf{p}_m\cdot\mathbf{r}_1} \phi_n(\mathbf{r}_1)\Bigr\vert^2.
\ee
which is proportional to the momentum distribution (at momentum $\mathbf{p}_m$) of the nucleus.

For the wavefunctions $\phi_n$, harmonic oscillator wavefunctions were used.  The oscillator length $b=\sqrt{\frac{\hbar}{\mu\omega}}$ was chosen so that the mean-square radius $\bar{R^2}$ as calculated using the density $\rho(\mathbf{r})=\sum_n \vert \phi_n(\mathbf{r}) \vert^2$ was equal to the mean-square radius $\bar{R^2}$ as calculated using the Woods-Saxon form of the nuclear number density:
\be
\rho(r)=\frac{\rho_0}{1+e^{\frac{r-R}{a}}}
\ee
where $R=1.1\;A^{1/3}\;fm$ and $a=0.56\;fm$; $\rho_0$ is determined by normalizing $\int d^3r \rho(r)$ to $A$.  The values obtained were $b=8.67 \;GeV^{-1}$ for $^{12}$C and  $b=10.48 \;GeV^{-1}$ for $^{40}$Ca.  The free-space cross-sections used in the calculations were $\sigma_{tot}^{pN}=40\;mb$, and $\sigma_{tot}^{V N}=25\;mb$~~\cite{anderson71}. 
\begin{figure}[tbp]
     \begin{center}
        \subfigure[VMD ]{%
            \label{fig:A=12rho1b}
             \includegraphics[width=0.5\textwidth,height=2in]{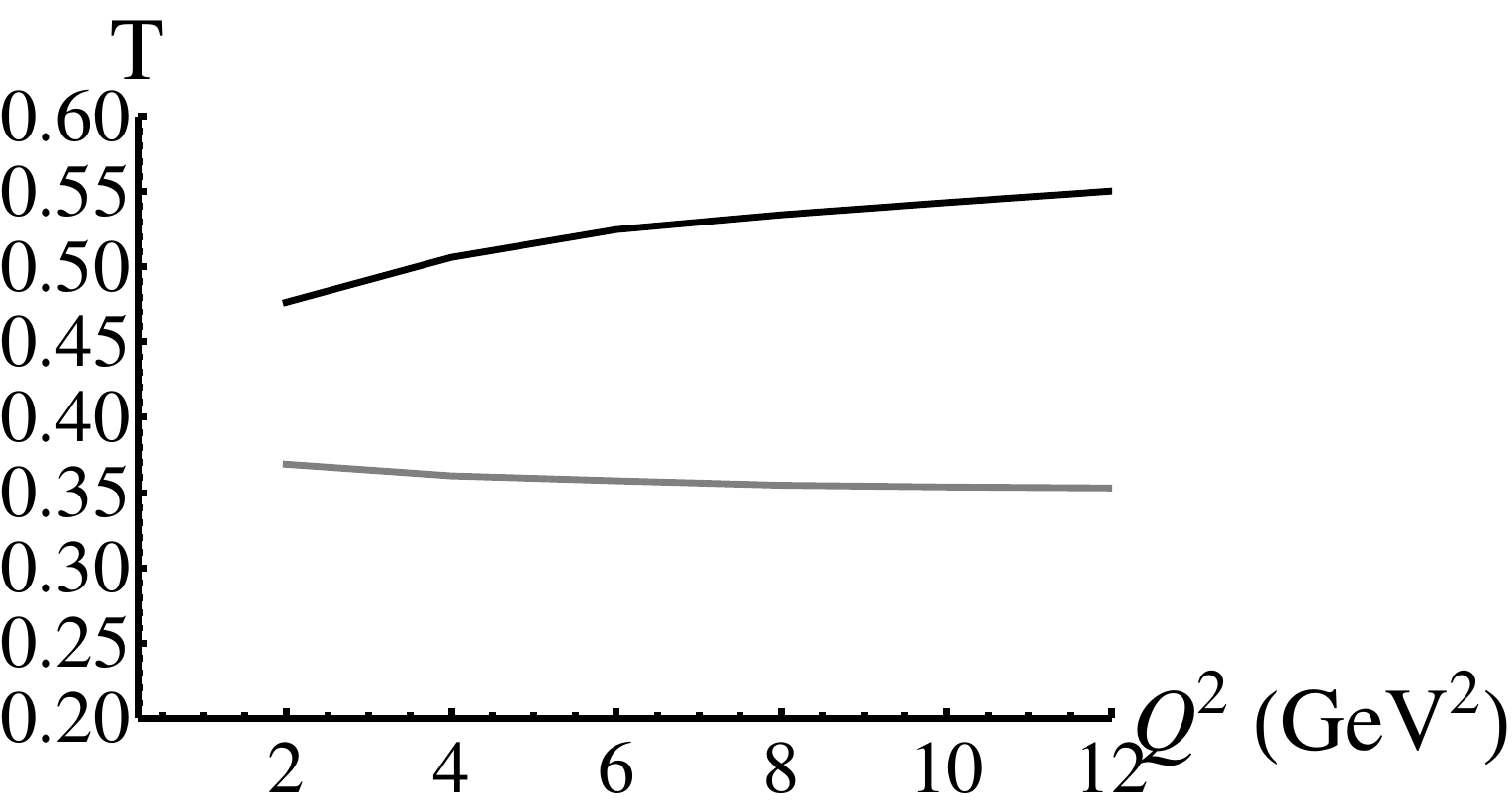}        
}%
        \subfigure[$b=7$ GeV$^{-2}$ ]{%
           \label{fig:A=12rho2}
             \includegraphics[width=0.5\textwidth,height=2in]{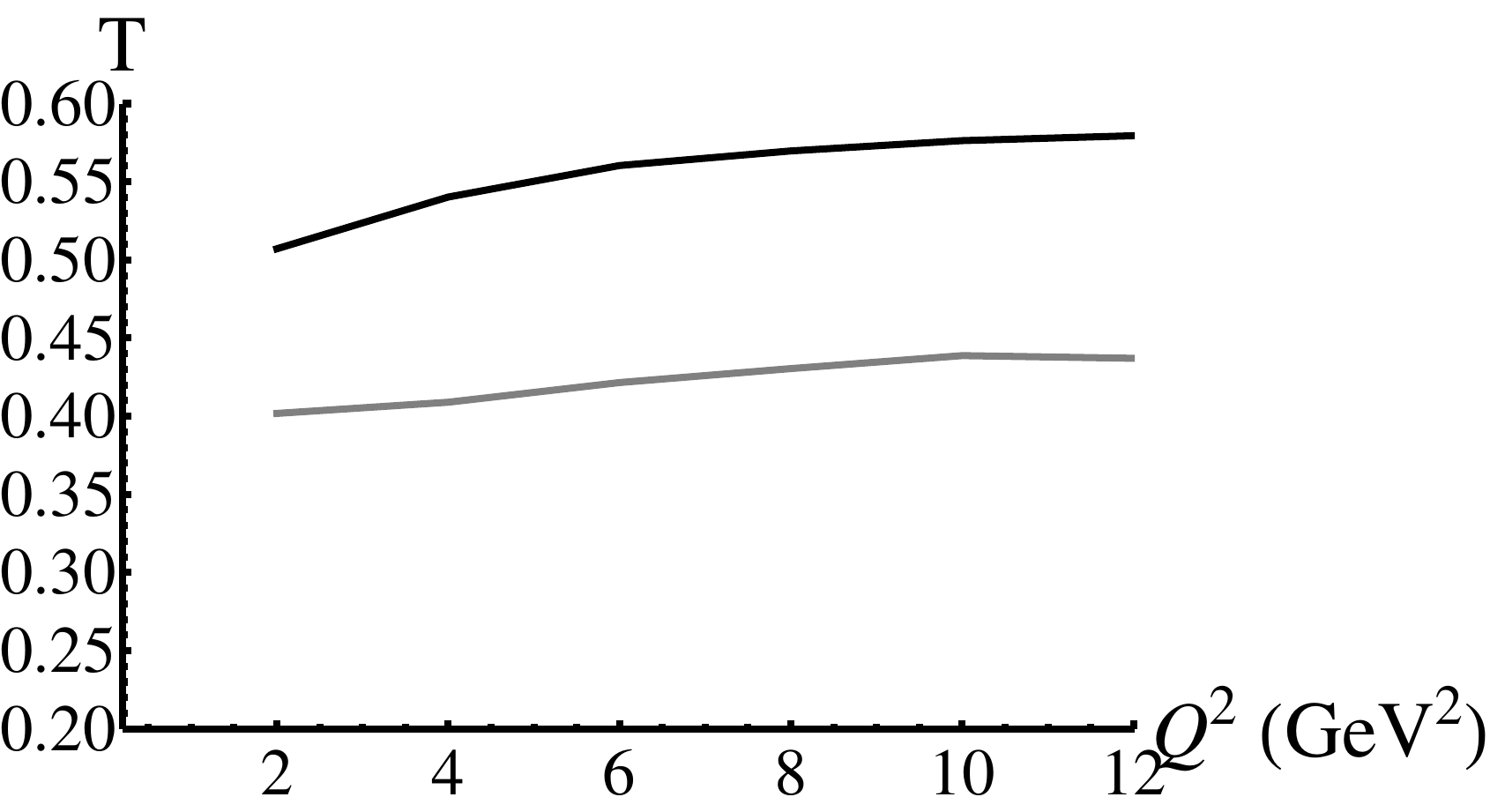}
        }\\ 

        \subfigure[ $b=8$ GeV$^{-2}$]{%
            \label{fig:A=12rho3}
		 \includegraphics[width=0.5\textwidth,height=2in]{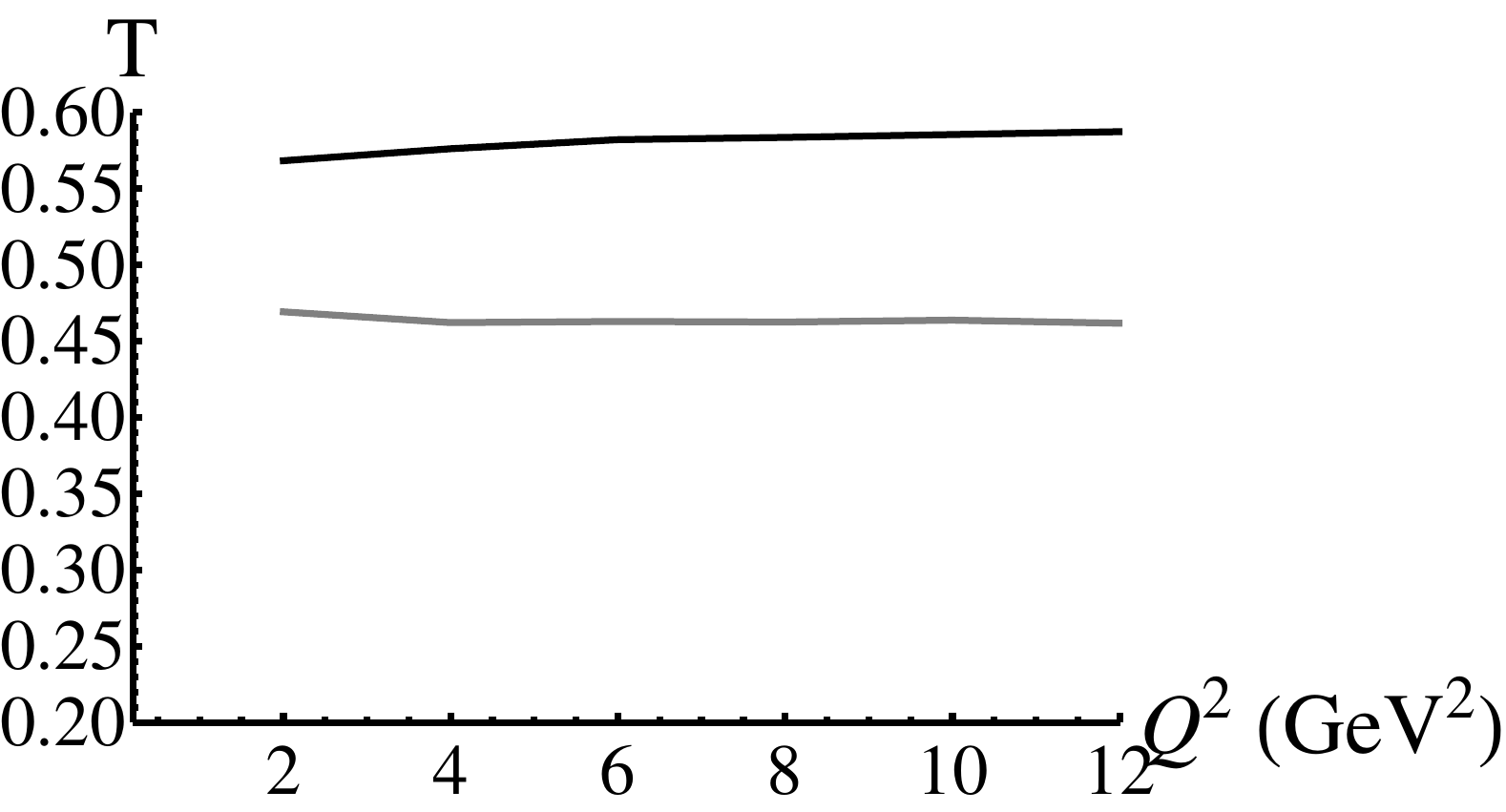}
        }%

    \end{center}
    \caption{%
        Transparency $T(\mathbf{p}_m)$ for $\mathbf{p}_m=0$ for $A=12$, $t=-2$ GeV$^2$, $l_c=5$ fm .  The bottom curves (gray) are the Glauber result; the top curves (black) are the CT result.  The value of the elastic $\rho$-nucleon $t$-slope parameter $b$ used in the calculation is indicated for each graph; VMD corresponds to $b_{\gamma V}=b$.
     }%
   \label{fig:pmzeroT}
\end{figure}

\begin{figure}[tbp]
     \begin{center}
        \subfigure[VMD ]{%
            \label{fig:A=40rho1b}
             \includegraphics[width=0.5\textwidth,height=2in]{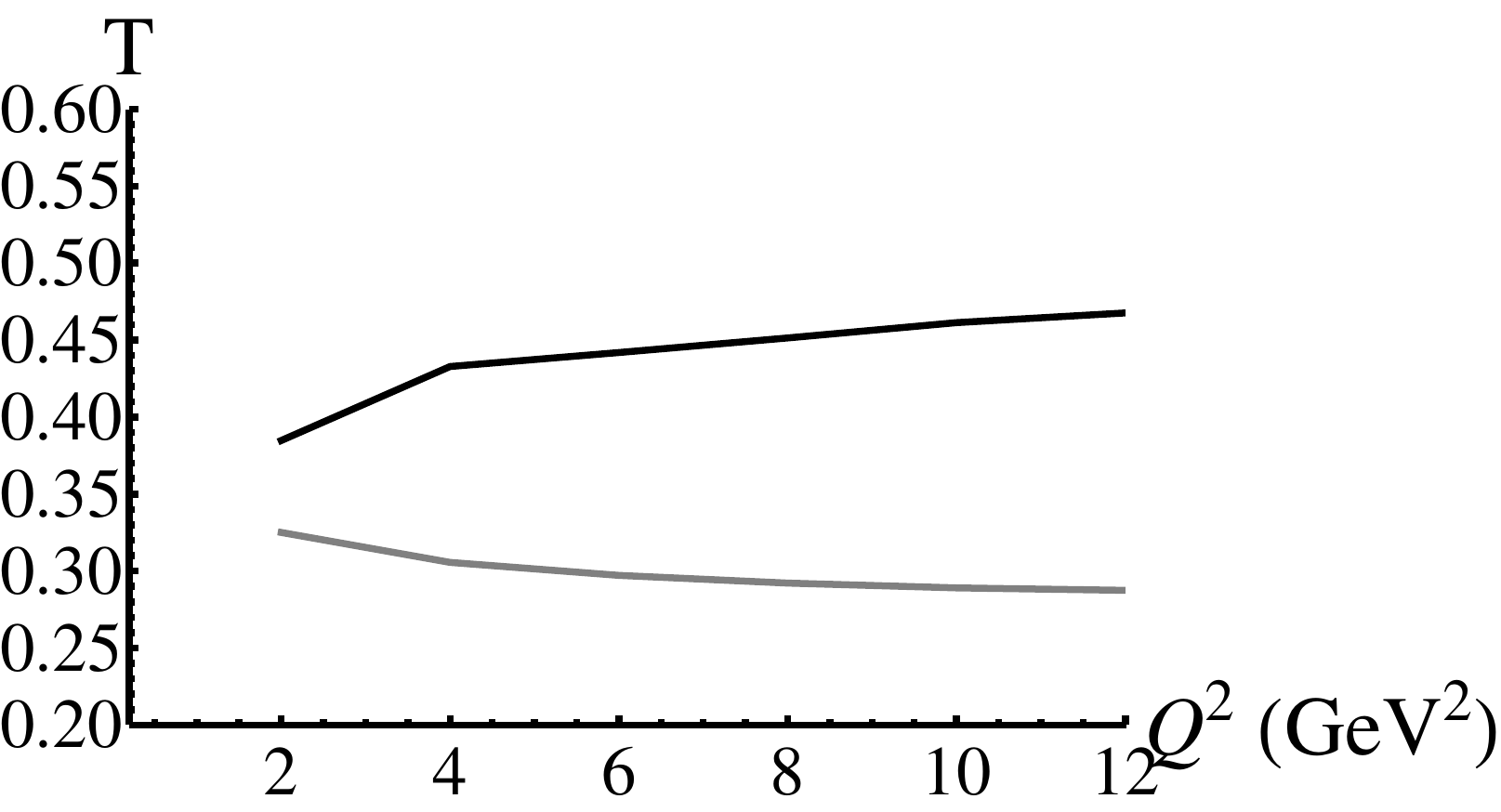}        
}%
        \subfigure[$b=7$ GeV$^{-2}$ ]{%
           \label{fig:A=40rho2}
            \includegraphics[width=0.5\textwidth,height=2in]{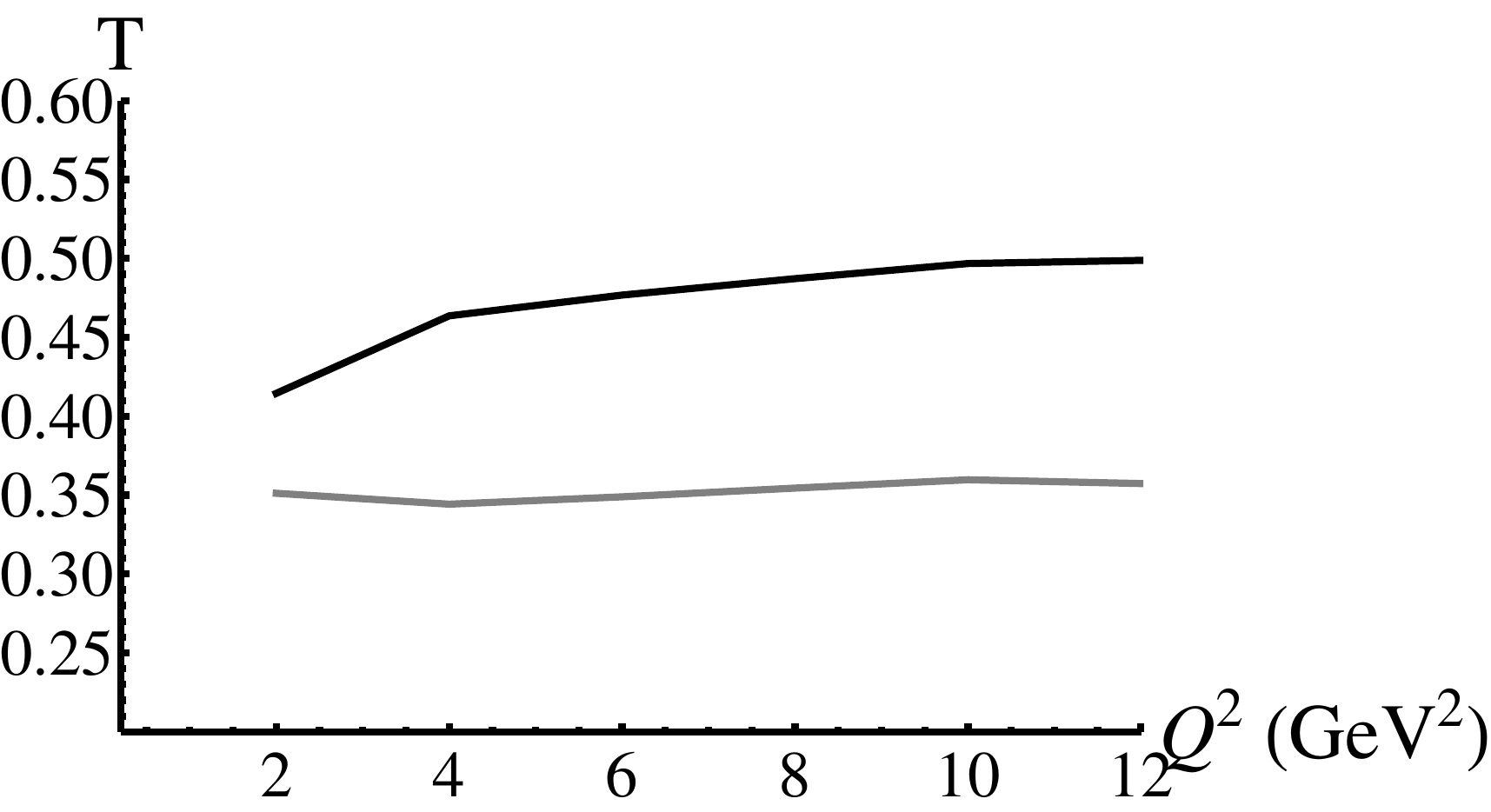}
        }\\ 

        \subfigure[ $b=8$ GeV$^{-2}$]{%
            \label{fig:A=40rho3}
		 \includegraphics[width=0.5\textwidth,height=2in]{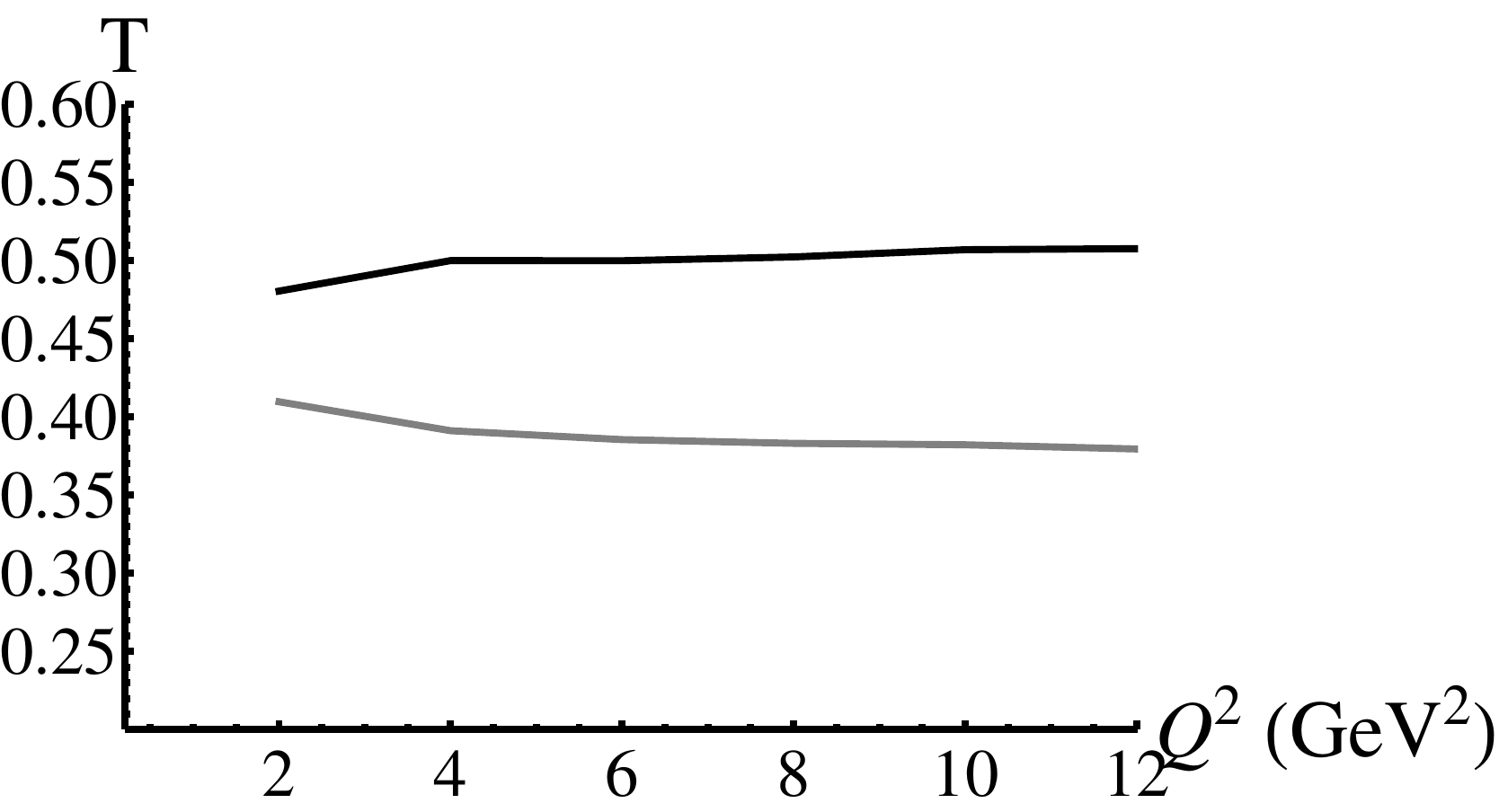}
        }%

    \end{center}
    \caption{%
        Transparency $T(\mathbf{p}_m)$ for $\mathbf{p}_m=0$ for $A=12$, $t=-2$ GeV$^2$, $l_c=5$ fm .  The bottom curves (gray) are the Glauber result; the top curves (black) are the CT result.  The value of the elastic $\rho$-nucleon $t$-slope parameter $b$ used in the calculation is indicated for each graph; VMD corresponds to $b_{\gamma V}=b$.
     }%
   \label{fig:pmzeroTA=40}
\end{figure}

The transparency $T$ was calculated for $^{12}C$ and $^{40}Ca$ at $\mathbf{p}_m=0$ for fixed $t$, and graphs of $T$ vs. $Q^2$ are shown in Figs. \ref{fig:pmzeroT} and \ref{fig:pmzeroTA=40}.  It is important to note that the transparency as a function of $t$ is calculated for fixed $\nu$ and $Q^2$, so that the coherence length $l_c$ is held constant.  If the coherence length varied, this could mimic Color Transparency because as $l_c$ gets smaller the attenuation due the Initial State Interaction of the vector meson (before the hard scatter) decreases since the vector meson propagates a smaller distance before undergoing the hard scatter; this would cause the value of the transparency $T$ to increase as $l_c$ decreases.  The production and elastic scattering amplitudes $f^{\gamma V}(\mathbf{q})$ and $f(\mathbf{q})$ in \eq{VresultCT} and \eq{denom} were taken to be of the form $f^{\gamma V}(\mathbf{q})=A_{\gamma V}e^{\frac{1}{2}b_{\gamma V}t}$ and $f^{\gamma V}(\mathbf{q})=Ae^{\frac{1}{2}bt}$ (where $t=-\mathbf{q}^2$) with the parameters  $A_{\gamma V}$, $b_{\gamma V}$, $A$ and $b$ taken from experimental data.  The t-slope $b$ for elastic $\rho$-nucleon scattering has been measured to be between $7$ and $8$ GeV$^{-2}$~\cite{anderson71}, while the t-slope for the production amplitude varies with $Q^2$.  The available electroproduction data~\cite{Chekanov:2007zr} are at higher virtual photon energies than are considered in this paper, but the values of $b_{\gamma V}(Q^2)$ measured in that experiment were what were used in our calculations.  Calculations were done for $b=7$ GeV$^{-2}$ and for $b=8$ GeV$^{-2}$ with $b_{\gamma V}$ depending on $Q^2$.  For comparison, calculations were also done assuming the validity of Vector Meson Dominance, in which case $b_{\gamma V}=b$ and the transparency $T(\mathbf{p}_m)$, \eq{transp}, is independent of the value of $b$ since both numerator and denominator are proportional to $e^{bt}$.

The expected properties of the transparency are evident in  Fig. \ref{fig:pmzeroT}.  For a given value of $Q^2$, the transparency (both Glauber and CT results) decreases with increasing $A$.    For a given $A$, as $Q^2$ increases the transparency in the CT case increases, which is also expected.  However, for the Glauber case, the behavior of $T$ as $Q^2$ varies is more sensitive to the values of $b$ and $b_{\gamma V}$ that are used.  Some of the dependence of $T$ on $Q^2$ is also due to the dependence of $\alpha_{\bfp} (\bfr)$ on kinematics through the relation \eq{pmdef}.

\section{Integrated transparency}
\label{sec:third}

The actual experimental situation corresponds to detection of the outgoing momentum corresponding to a range of values of the missing momentum $\mathbf{p}_m$.  The integrated transparency is 
\be
\label{intT1}
T_{\cal{D}}=\frac{\int_{\cal{D}}d^3p_m\frac{d\sigma}{dE'd\Omega'  d\Omega_p}}{\int_{\cal{D}}d^3p_m \frac{d\sigma_{PWIA}}{dE'd\Omega'  d\Omega_p}}=\frac{ \sum_{n=1}^A\int_{\cal{D}}d^3p_m \Bigl\vert F^{(n)}(\mathbf{p}_m)\Bigr\vert^2}{ \sum_{n=1}^A\int_{\cal{D}}d^3p_m \Bigl\vert F^{(n)}(\mathbf{p}_m)\Bigr\vert_{PWIA}^2}.
\ee 
In the impulse approximation, the missing momentum $\mathbf{p}_m$ is equal to the negative of the momentum that the struck proton had inside the nucleus before the collision.  Therefore the amplitude $F_{fi}^{(n)}$ as a function of $\mathbf{p}_m$ should be negligible for $p_m>300\;MeV$ or so, since the momentum of the nucleons in the nucleus cannot be much larger than this.

If we integrate over $p_m$ up to $p_{max}\simeq 300\; MeV$, we may set $\mathbf{p}=\mathbf{q}+\mathbf{p}_m\simeq\mathbf{q}$ in $\alpha_{\mathbf{p}}$, since for the kinematics of interest we have $p,\;q\gg 300\;MeV$.  Then assuming that $F_{fi}^{(n)}$     is zero for $\mathbf{p}_m>p_{max}$, we may extend the upper limit of integration in \eq{intT1} to infinity, $p_{max}\to\infty$.  For the denominator we obtain simply $(2\pi)^3\;A\;\vert f^{\gamma V}(\mathbf{q})\vert^2$.       For the numerator we obtain three terms:
\be
\label{eq:numerator}
(2\pi)^3\vert f^{\gamma V}(\mathbf{q})\vert^2  \int d^2s_1  d z_1 \rho(\mathbf{r}_1) e^{-\alpha_p(\mathbf{r}_1)} (h_1(\mathbf{r}_1)+h_2(\mathbf{r}_1)+h_3(\mathbf{r}_1))
\ee
where
\begin{flalign}
&h_1(\mathbf{r}_1) =   e^{-\alpha_V(\mathbf{s}_1,z_1)},&
\end{flalign}
\be
h_2(\mathbf{r}_1)=\frac{4\pi}{ik} \frac{f^{\gamma V}(\mathbf{q})f^{\gamma V}(0)}{\vert f^{\gamma V}(\mathbf{q})\vert^2}  e^{-\frac{1}{2}\alpha_V(\mathbf{s}_1,z_1)}\int_{-\infty}^{z_1} dz_2 \rho(\mathbf{s}_1,z_2) e^{-\frac{1}{2}\alpha_V(\mathbf{s}_1,z_2)}\cos{q_L(z_1-z_2)  f^{PLC}(z_1,z_2,\mathbf{q})},
\ee
and
\be
\begin{split}
h_3(\mathbf{r}_1)= \Bigl(\frac{2\pi}{k}\Bigr)^2 \frac{\vert f^{\gamma V}(0)\vert^2}{\vert f^{\gamma V}(\mathbf{q})\vert^2}        \int_{-\infty}^{z_1} dz_2 \int_{-\infty}^{z_1} dz_3  & \rho(\mathbf{s}_1,z_2) \rho(\mathbf{s}_1,z_3)  e^{-\frac{1}{2}\alpha_V(\mathbf{s}_1,z_2)} e^{-\frac{1}{2}\alpha_V(\mathbf{s}_1,z_3)}  \cos{q_L(z_2-z_3)}\\
& \times f^{PLC}(z_1,z_2,\mathbf{q}) f^{*PLC}(z_1,z_3,\mathbf{q})  
.\end{split}
\ee
Thus we have for the integrated transparency
\be
\label{eq:integratedT}
T_{{\cal D}}=\frac{1}{A}\int d^2s_1  d z_1 \rho(\mathbf{r}_1) e^{-\alpha_p(\mathbf{r}_1)} (h_1(\mathbf{r}_1)+h_2(\mathbf{r}_1)+h_3(\mathbf{r}_1)).
\ee
This simplifies somewhat if we assume the validity of Vector Meson Dominance for the relation between the free-space production amplitude $f^{\gamma V}(\mathbf{q})$ and the free-space elastic scattering amplitude $f(\mathbf{q})$ (which appears inside $f^{PLC}$; see Eq.\ref{eq:fplc}).  Assuming that the high-energy amplitudes are purely imaginary, use of the optical theorem then gives:

\be
\label{eq:h2}
\begin{split}
h_2(\mathbf{r}_1)=- \frac{\sigma^{tot}_{VN}}{G_V(t)} e^{-\frac{1}{2}\alpha_V(\mathbf{s}_1,z_1)}\int_{-\infty}^{z_1} dz_2 \rho(\mathbf{s}_1,z_2)& e^{-\frac{1}{2}\alpha_V(\mathbf{s}_1,z_2)}\cos{q_L(z_1-z_2)}\\
& \times h(z_1-z_2)G_V\Bigl(t\;h(z_1-z_2)\Bigr)
\end{split}
\ee
\be
\label{eq:h3}
\begin{split}
h_3(\mathbf{r}_1)= \frac{1}{4}\Bigl(\frac{\sigma^{tot}_{VN}}{G_V(t)} \Bigr)^2       \int_{-\infty}^{z_1} &dz_2 \int_{-\infty}^{z_1} dz_3   \rho(\mathbf{s}_1,z_2) \rho(\mathbf{s}_1,z_3)  e^{-\frac{1}{2}\alpha_V(\mathbf{s}_1,z_2)} e^{-\frac{1}{2}\alpha_V(\mathbf{s}_1,z_3)}  \cos{q_L(z_2-z_3)}\\
& \times h(z_1-z_2) h(z_1-z_3) G_V\Bigl(t\;h(z_1-z_2)\Bigr) G_V\Bigl(t\;h(z_1-z_3)\Bigr)
\end{split}
\ee
where
\be
h(z)\equiv \frac{\sigma^{eff}_{VN}(z,Q^2)}{\sigma^{tot}_{VN}}=\Biggl[ \theta(l_h-z)\; \Bigl[\frac{z}{l_h}+\frac{n^2\langle k_t^2\rangle}{Q^2}\Bigl (1-\frac{z}{l_h}\Bigr) \Bigr]  +\theta(z-l_h)\Biggr].
\ee
The form factor $G_V$ used in evaluating \eq{eq:integratedT}  was taken to be the same form factor as for the pion: \be
G_V(t)=\frac{1}{1-t/0.59},
\ee
for $t$ in GeV$^2$.

The 3 terms of \eq{eq:numerator} or \eq{eq:integratedT} are represented pictorially by the same diagrams as in Fig. \ref{fig:pictorialrep}.  The term with $h_1$ is the square of the diagram in Fig. \ref{fig:pictorialrep1} and represents  incoherent production from nucleon 1; the term with $h_2$ represents interference between the diagrams of Fig. \ref{fig:pictorialrep1} and Fig. \ref{fig:pictorialrep2}, with interference between production on nucleon 1 and nucleon 2; and the term with $h_3$ is the square of the diagram in Fig. \ref{fig:pictorialrep2}, which represents interference between production on nucleon 2 and production on a different nucleon 3, with incoherent scattering from nucleon 1.

\begin{figure}[tbp]
     \begin{center}

        \subfigure[ VMD]{%
            \label{fig:A=56rho1}
             \includegraphics[width=0.7\textwidth,height=2.2in]{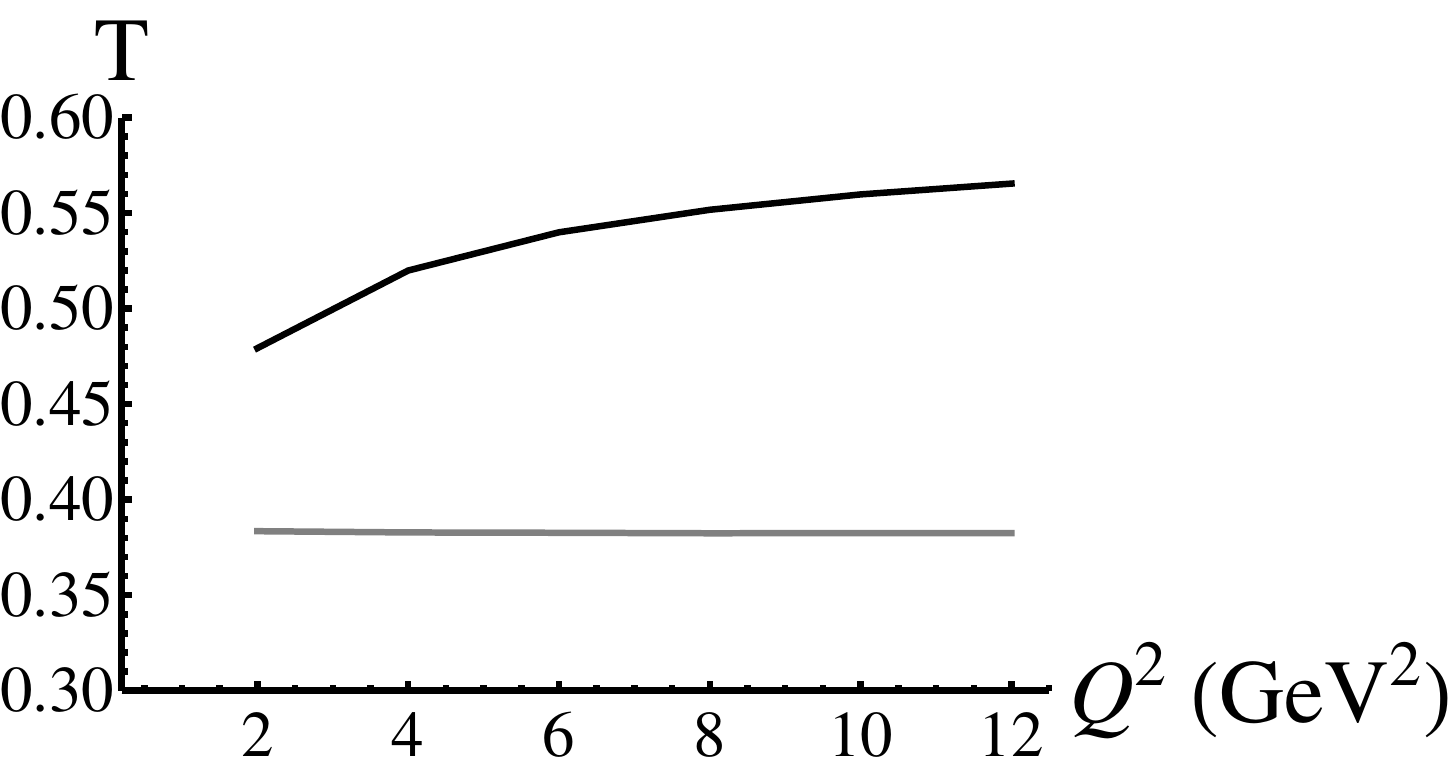}
        }%
    \hspace{0.5in}
        \subfigure[  $b=7$ GeV$^{-2}$]{%
           \label{fig:A=56rho2}
           \includegraphics[width=0.7\textwidth,height=2.2in]{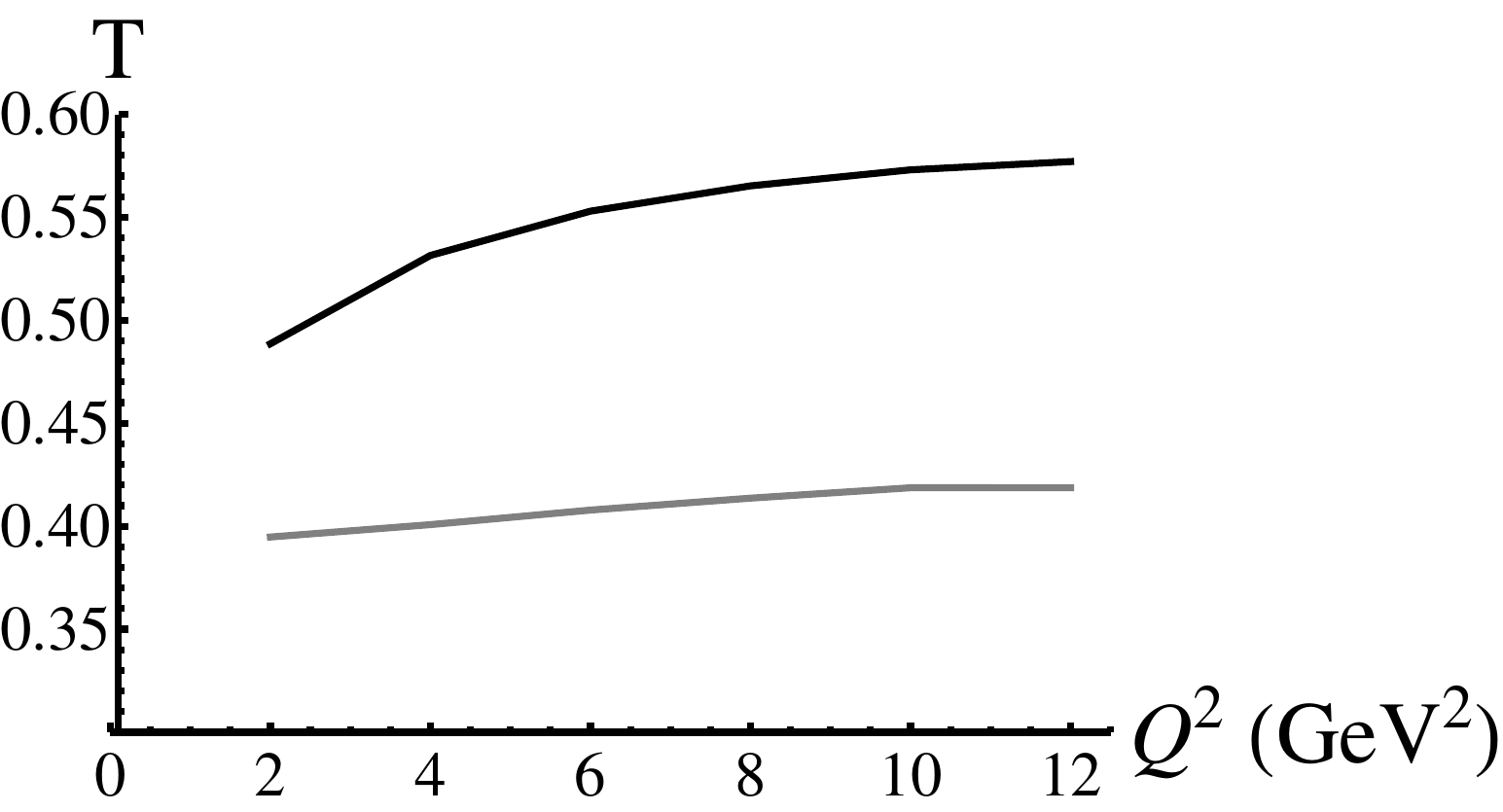}
        }\\ 
        \subfigure[ $b=8$ GeV$^{-2}$]{%
            \label{fig:A=12rho3}
		\includegraphics[width=0.7\textwidth,height=2.2in]{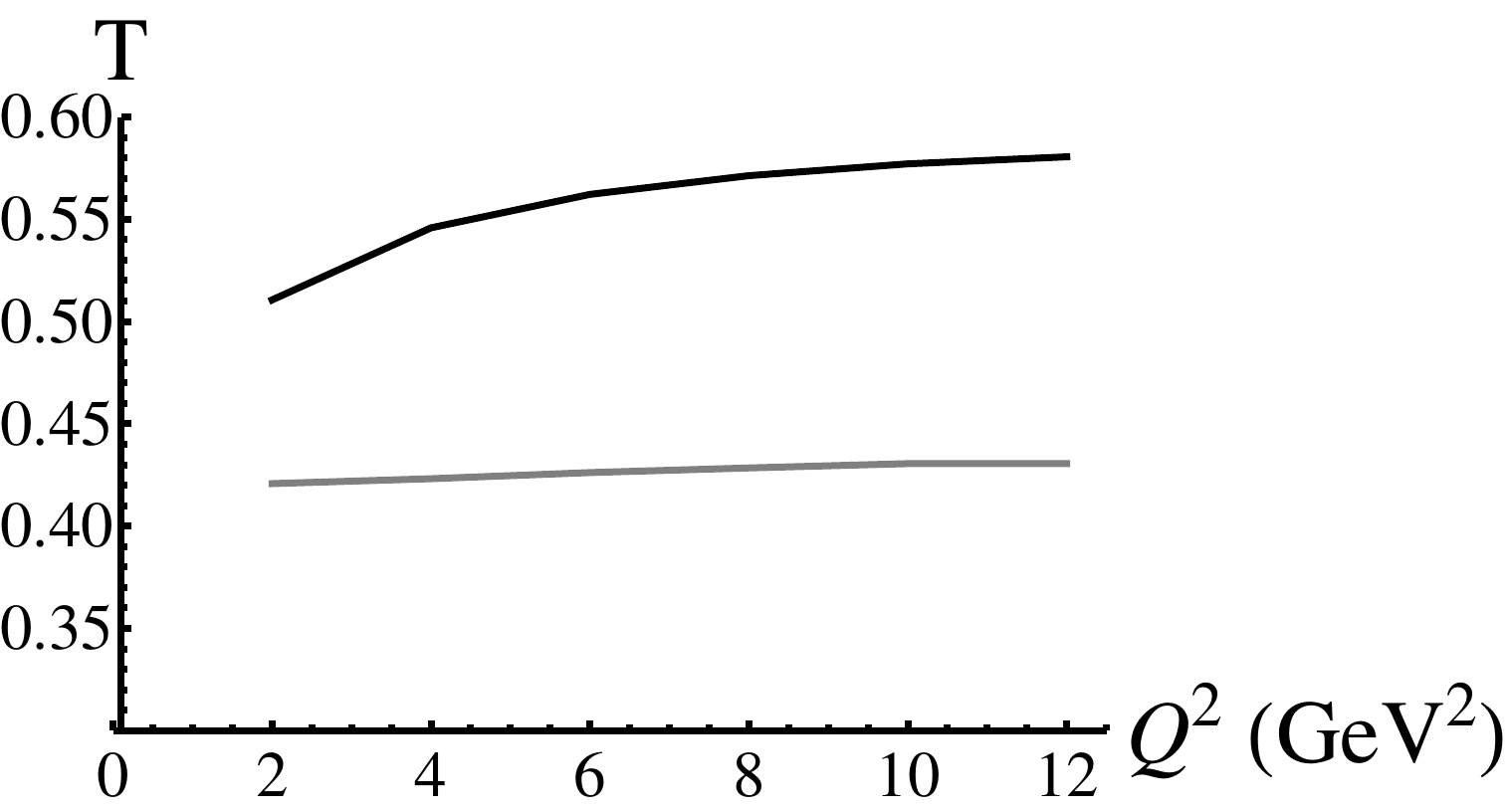}
        }%
    \end{center}
    \caption{%
        Integrated transparency $T$ for  $A=12$, $t=-2$ GeV$^2$, $l_c=2$ fm.  
        The bottom curves (gray) are the Glauber result; the top curves (black) are the CT result.  The value of the elastic $\rho$-nucleon $t$-slope parameter $b$ used in the calculation is indicated for each graph; VMD corresponds to $b_{\gamma V}=b$.
     }%
   \label{fig:intTA=12}
\end{figure}

\begin{figure}[tbp]
     \begin{center}

        \subfigure[ VMD]{%
            \label{fig:A=56rho1}
             \includegraphics[width=0.7\textwidth,height=2.2in]{A=12Lc=2.pdf}
        }%
    \hspace{0.5in}
        \subfigure[  $b=7$ GeV$^{-2}$]{%
           \label{fig:A=56rho2}
           \includegraphics[width=0.7\textwidth,height=2.2in]{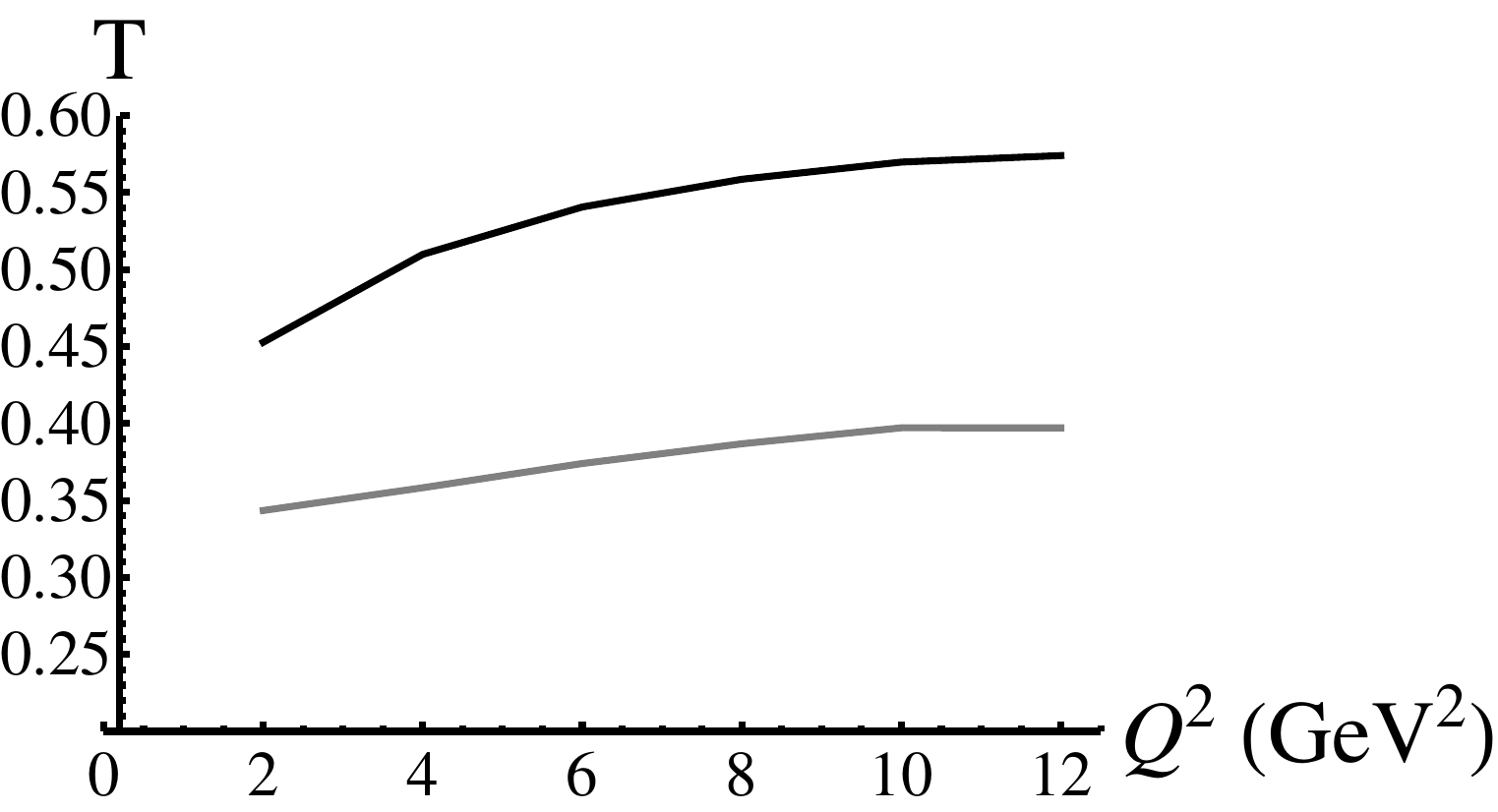}
        }\\ 
        \subfigure[ $b=8$ GeV$^{-2}$]{%
            \label{fig:A=12rho3}
		 \includegraphics[width=0.7\textwidth,height=2.2in]{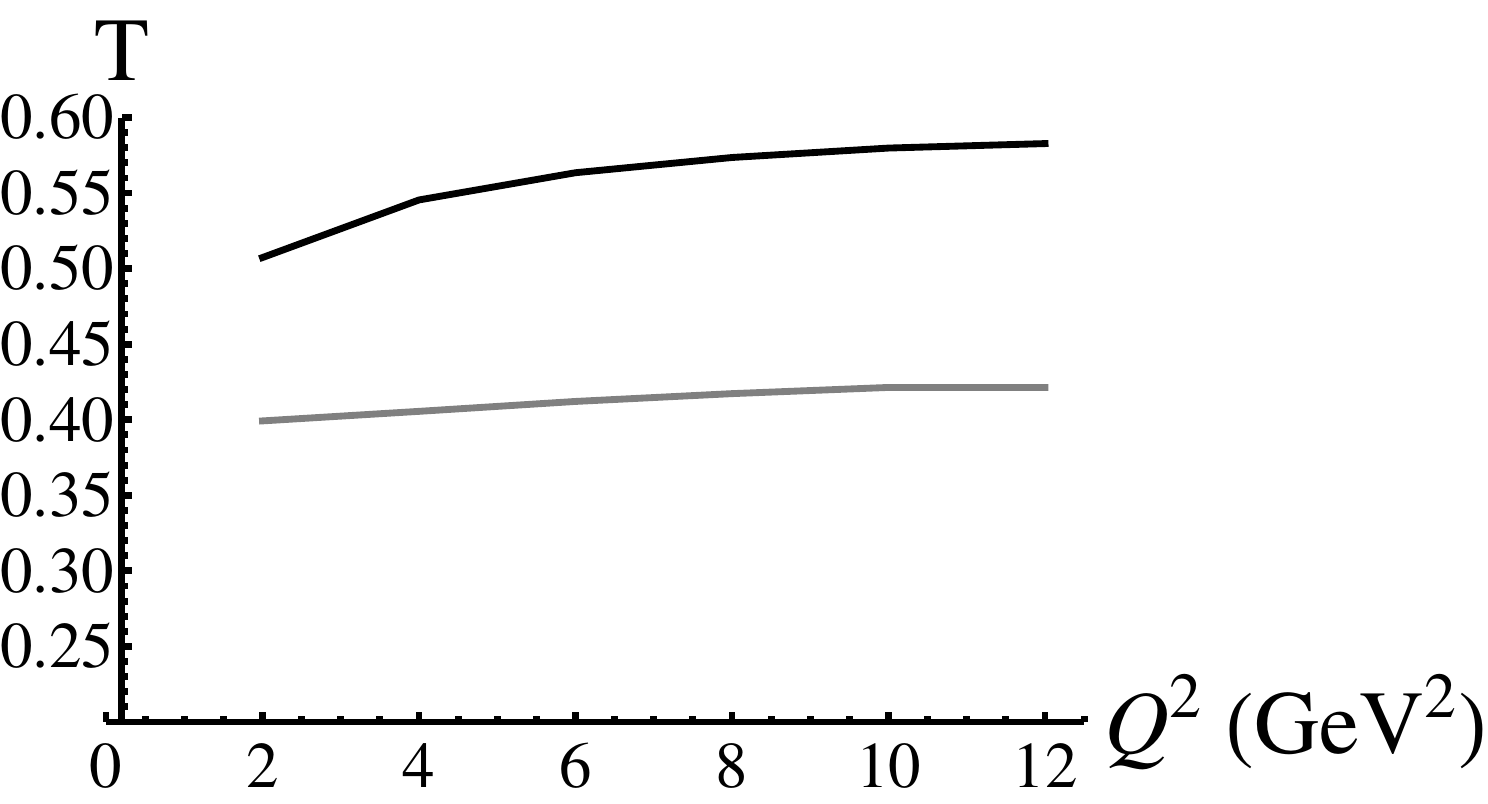}
        }%
    \end{center}
    \caption{%
        Integrated transparency $T$ for  $A=12$, $t=-2$ GeV$^2$, $l_c=5$ fm.  
        The bottom curves (gray) are the Glauber result; the top curves (black) are the CT result.  The value of the elastic $\rho$-nucleon $t$-slope parameter $b$ used in the calculation is indicated for each graph; VMD corresponds to $b_{\gamma V}=b$.
     }%
   \label{fig:intTA=12b}
\end{figure}

\begin{figure}[tbp]
     \begin{center}

        \subfigure[ VMD]{%
            \label{fig:A=56rho1}
             \includegraphics[width=0.7\textwidth,height=2.2in]{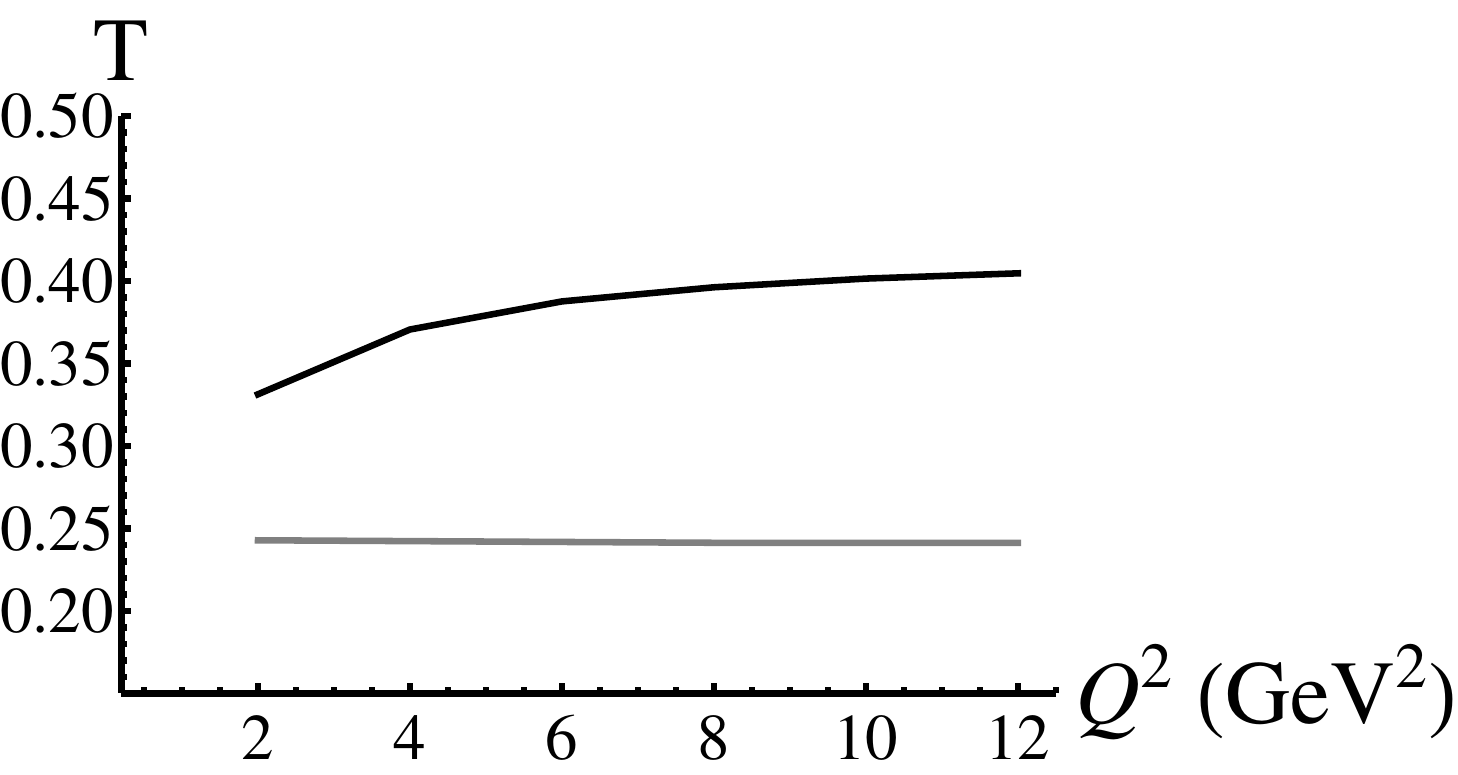}
        }%
    \hspace{0.5in}
        \subfigure[  $b=7$ GeV$^{-2}$]{%
           \label{fig:A=56rho2}
           \includegraphics[width=0.7\textwidth,height=2.2in]{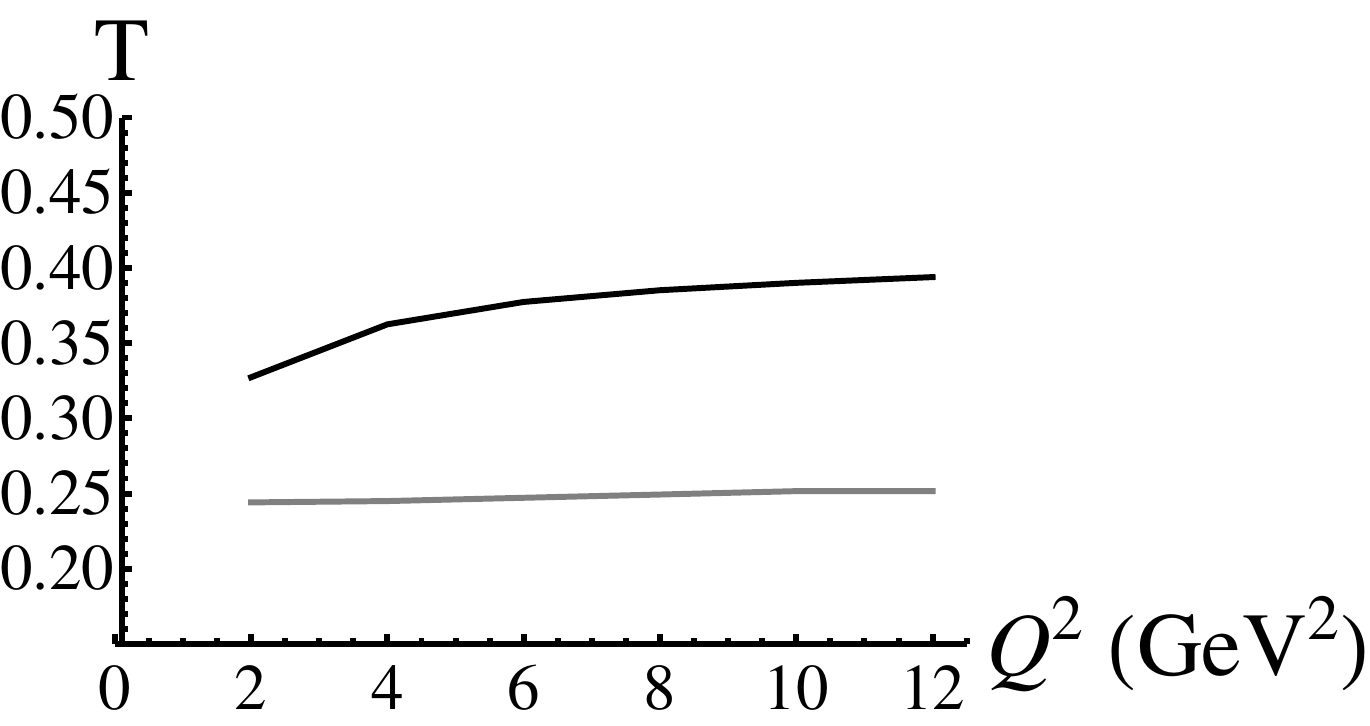}
        }\\ 
        \subfigure[ $b=8$ GeV$^{-2}$]{%
            \label{fig:A=12rho3}
		\includegraphics[width=0.7\textwidth,height=2.2in]{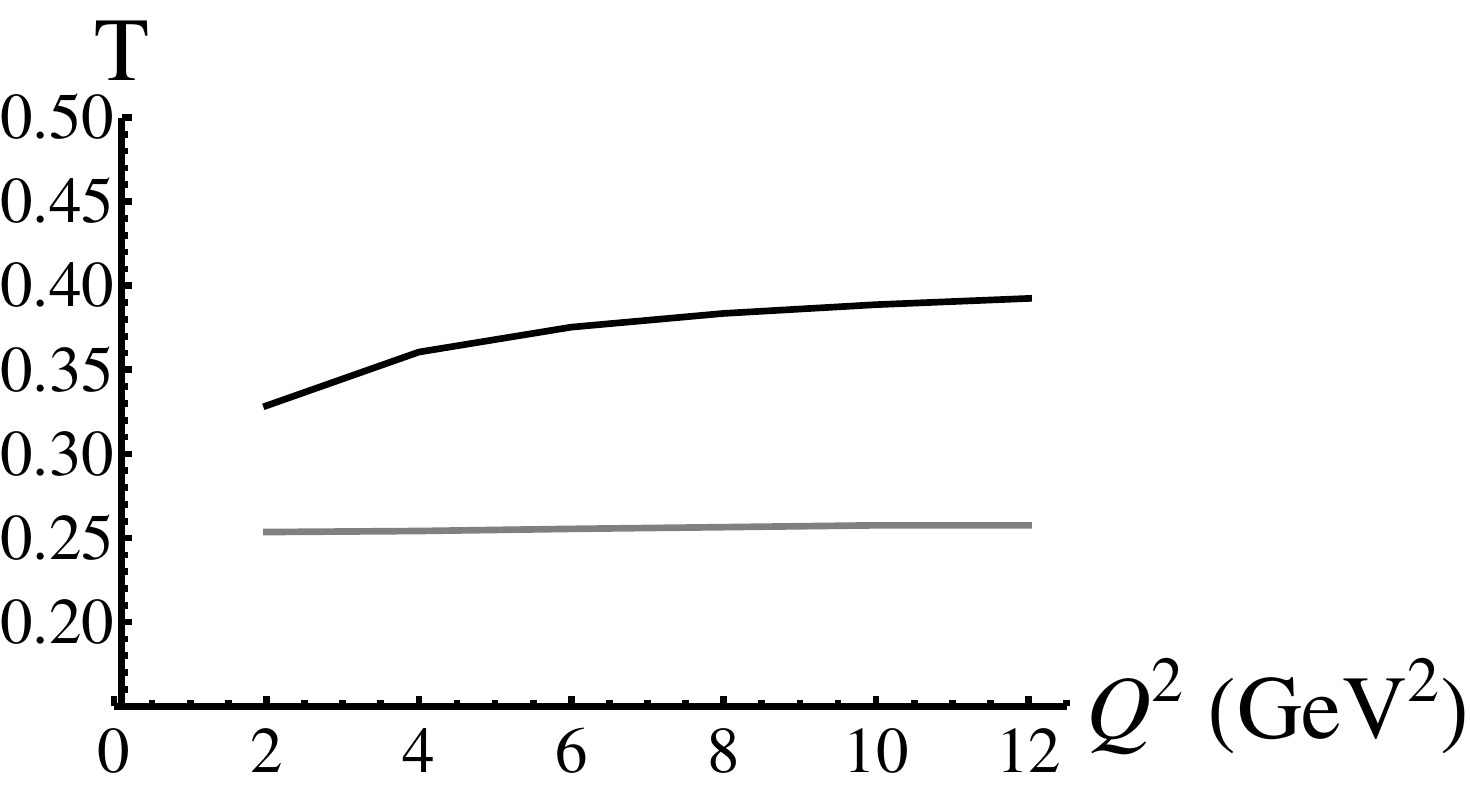}
        }%
    \end{center}
    \caption{%
        Integrated transparency $T$ for  $A=40$, $t=-2$ GeV$^2$, $l_c=2$ fm.  
        The bottom curves (gray) are the Glauber result; the top curves (black) are the CT result.  The value of the elastic $\rho$-nucleon $t$-slope parameter $b$ used in the calculation is indicated for each graph; VMD corresponds to $b_{\gamma V}=b$.
     }%
   \label{fig:intTA=40}
\end{figure}

\begin{figure}[tbp]
     \begin{center}

        \subfigure[ VMD]{%
            \label{fig:A=56rho1}
             \includegraphics[width=0.7\textwidth,height=2.2in]{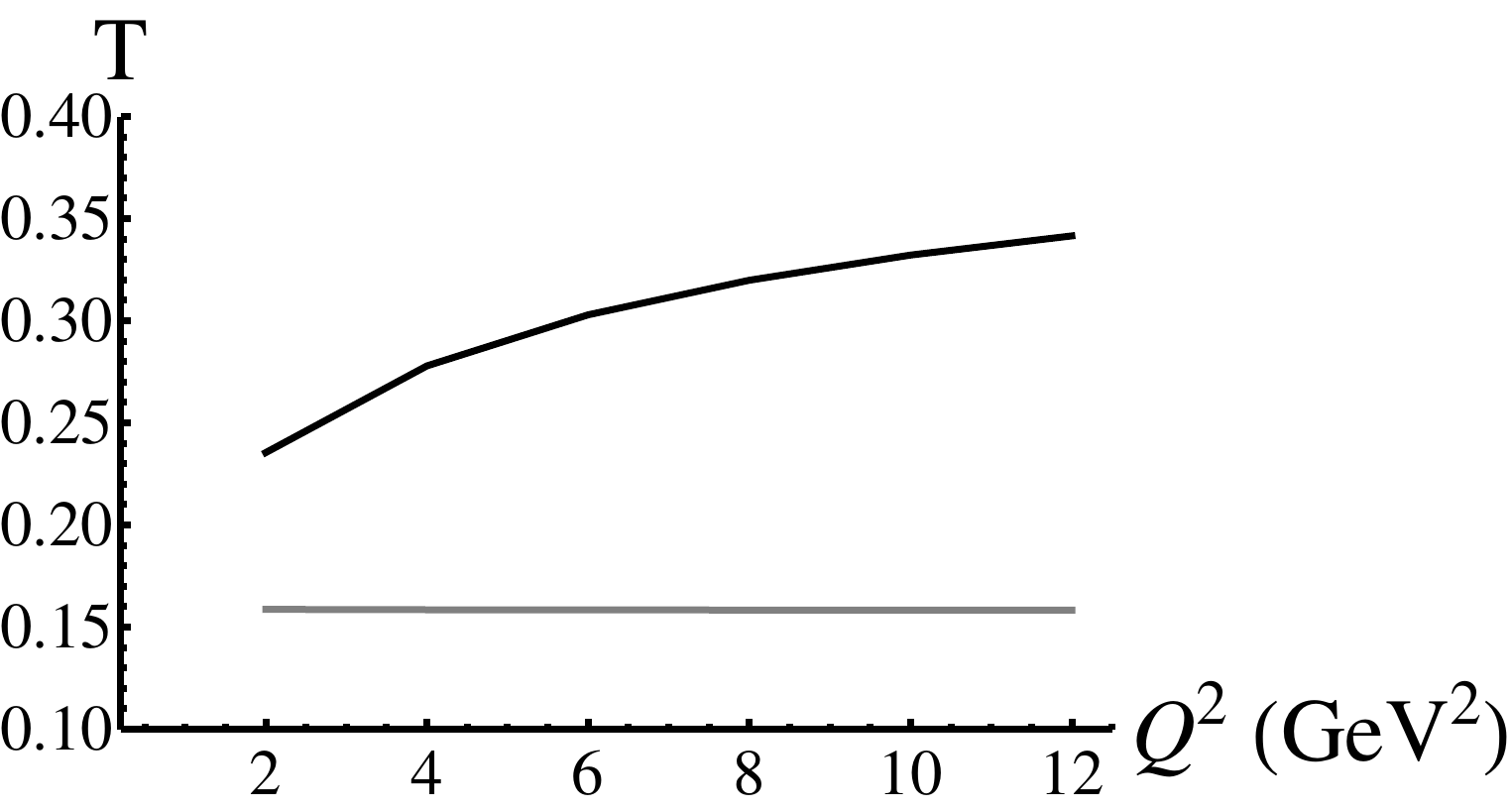}
        }%
    \hspace{0.5in}
        \subfigure[  $b=7$ GeV$^{-2}$]{%
           \label{fig:A=56rho2}
           \includegraphics[width=0.7\textwidth,height=2.2in]{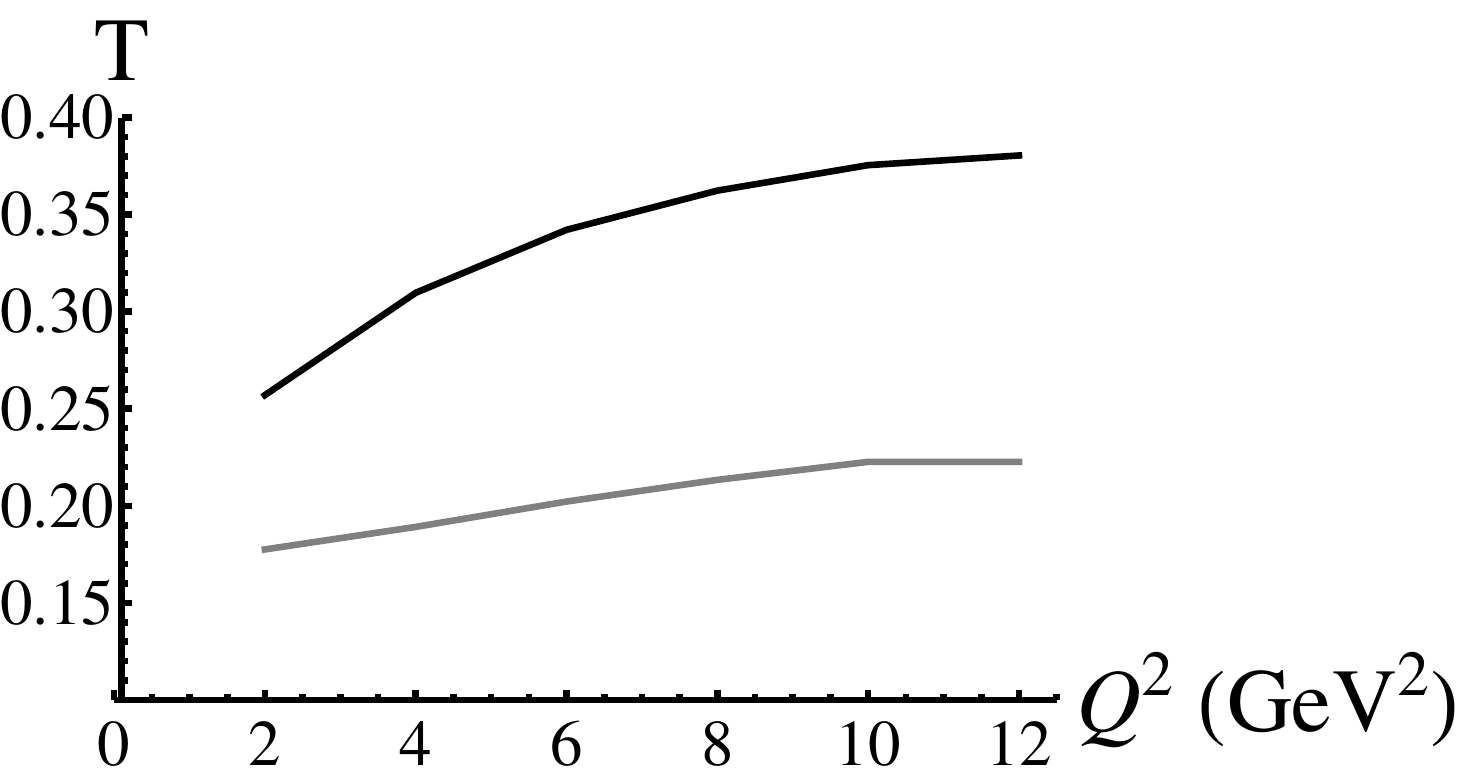}
        }\\ 
        \subfigure[ $b=8$ GeV$^{-2}$]{%
            \label{fig:A=12rho3}
		 \includegraphics[width=0.7\textwidth,height=2.2in]{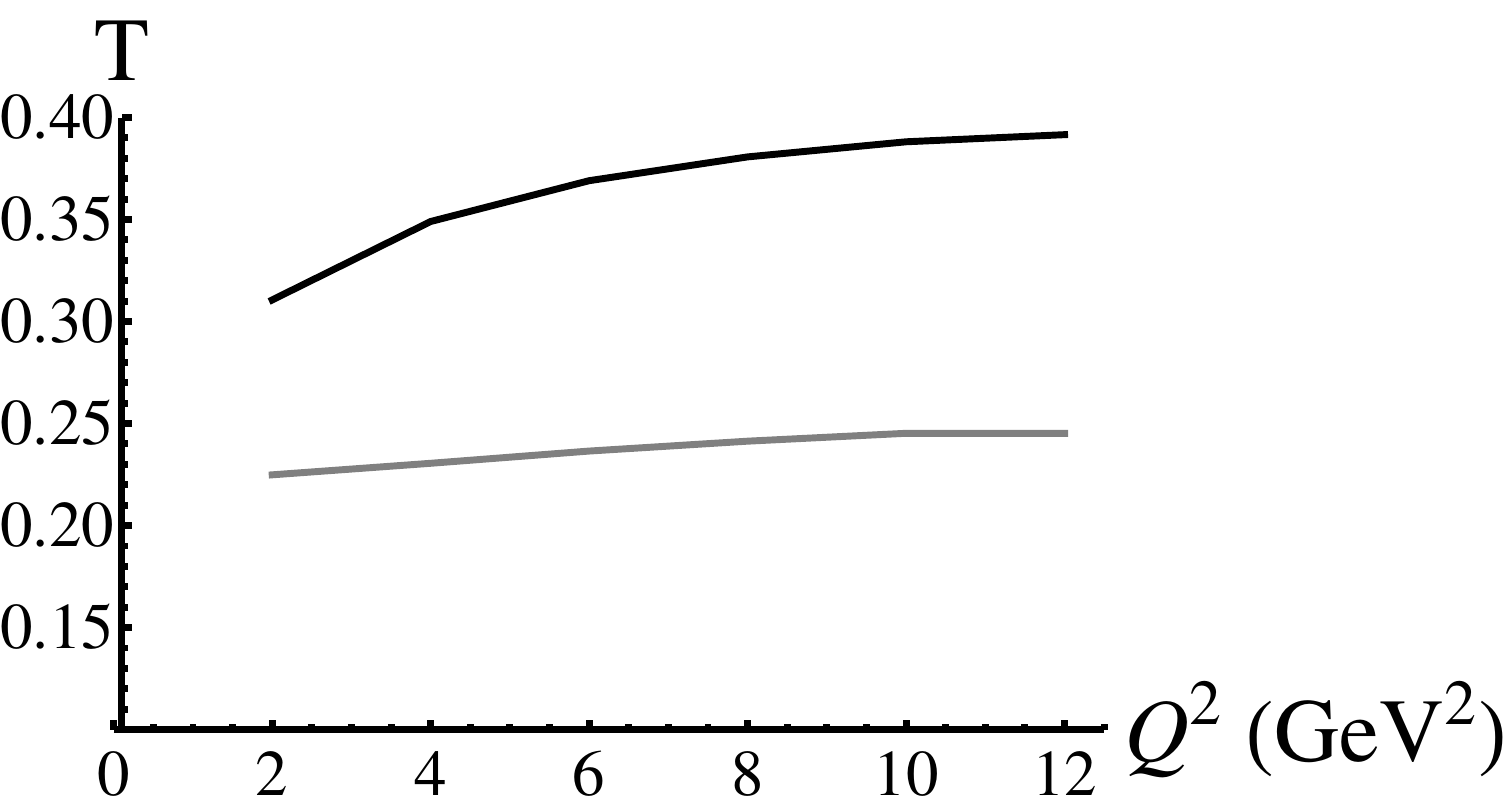}
        }%
    \end{center}
    \caption{%
        Integrated transparency $T$ for  $A=40$, $t=-2$ GeV$^2$, $l_c=5$ fm.  
        The bottom curves (gray) are the Glauber result; the top curves (black) are the CT result.  The value of the elastic $\rho$-nucleon $t$-slope parameter $b$ used in the calculation is indicated for each graph; VMD corresponds to $b_{\gamma V}=b$.
     }%
   \label{fig:intTA=40b}
\end{figure}

\begin{figure}[tbp]
     \begin{center}
        \subfigure[ $A=12$, $Q^2=0.5$ GeV$^2$, $l_c=5$ fm]{%
            \label{fig:A=56rho1}
             \includegraphics[width=0.7\textwidth,height=2.2in]{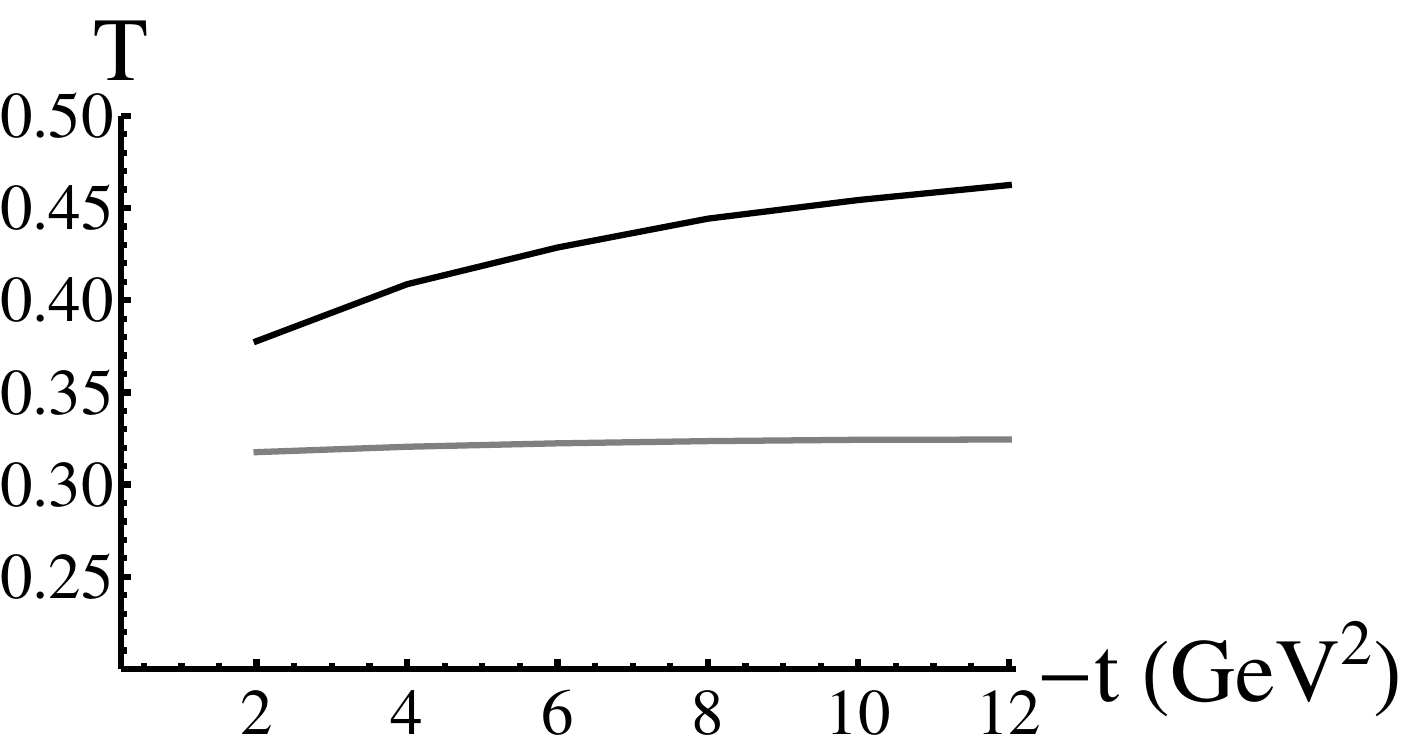}
        }%
 \hspace{0.5in}
        \subfigure[  $A=40$, $Q^2=0.5$ GeV$^2$, $l_c=5$ fm]{%
           \label{fig:A=56rho2}
            \includegraphics[width=0.7\textwidth,height=2.2in]{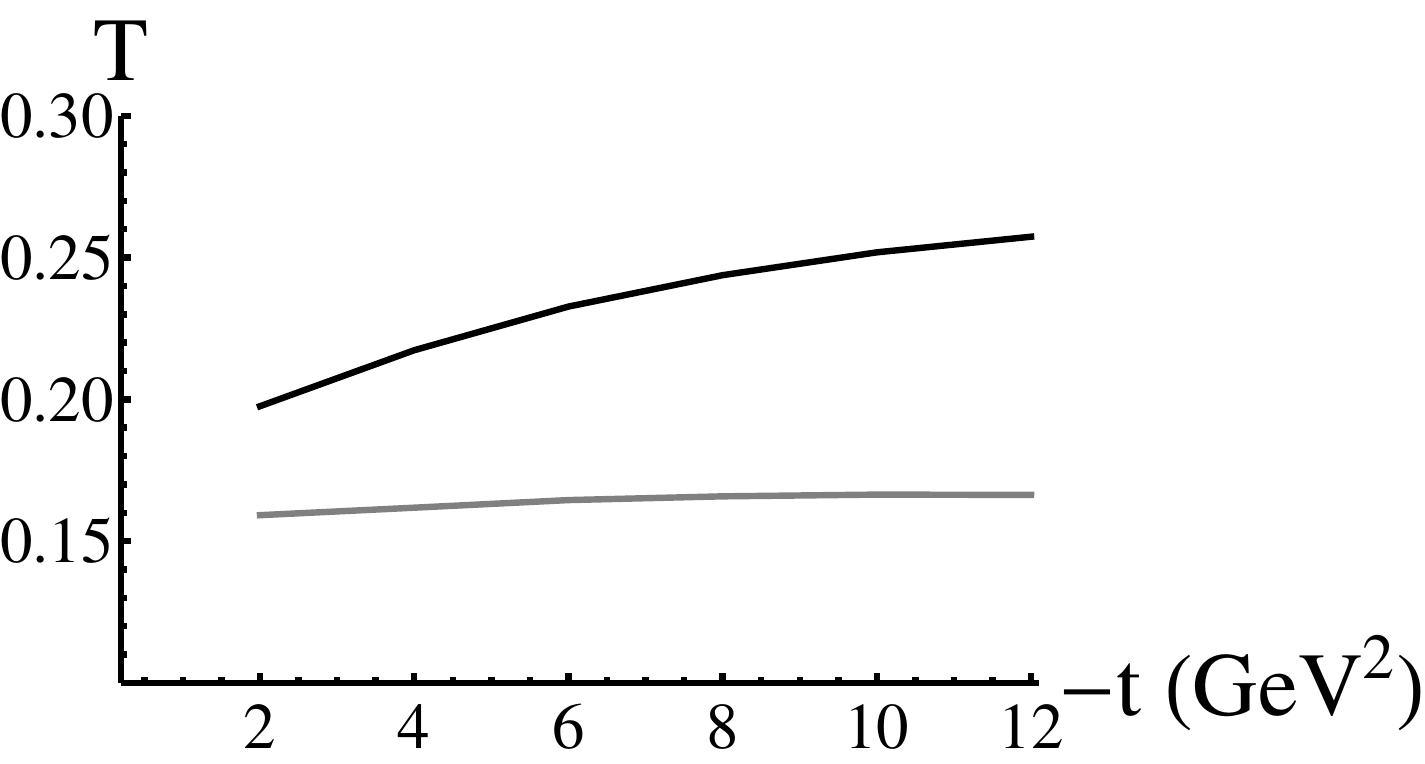}
        }\\ 
%
    \end{center}
    \caption{%
        Integrated transparency $T$ for fixed $Q^2$ and $l_c$ and varying $t$.  The bottom curves (gray) are the Glauber result; the top curves (black) are the CT result.
     }%
   \label{fig:intTfixedQ}
\end{figure}

The integrated transparency was calculated for $A=12$ and $A=40$, for a range of values of $t$ and $Q^2$.  In Figs. \ref{fig:intTA=12} - \ref{fig:intTA=40b}, the transparency is shown  for fixed $t$ as a function of $Q^2$, for two different values of the coherence length.  The same values of $b$ and $b_{\gamma V}$ were used as for the $T(\mathbf{p}_m=0)$ calculation; VMD corresponds to $b=b_{\gamma V}$.

The same overall features of the graphs are present as were seen for the $\mathbf{p}_m=0$ transparency.  In addition, here one can see that for a given $A$ and $Q^2$, the transparency increases as the coherence length $l_c$ decreases, which agrees with expectations.  For the whole range of $Q^2$ from $2$ to $12$ GeV$^2$, the difference between the CT transparency and the Glauber transparency is significant.  For the higher values of $Q^2$, the CT value is of the order of $1.5$ times as large as the Glauber transparency, for $A=12$, and $2$ times as large as the Glauber transparency for $A=40$. The integrated transparency is significantly smaller than the values for ${\bf p}_m=0$. This is a relevant feature for experimentalists to note.

In Fig. \ref{fig:intTfixedQ}, the transparency is shown for fixed $Q^2$ as a function of $t$.  In that figure, $Q^2=0.5$ GeV$^2$, which is small enough that for the rescattering terms (Eqs. \ref{eq:h2} and \ref{eq:h3}) the produced $q\bar{q}$ (at either $z_2$ or $z_3$) is a normal $\rho$-meson.  Thus no Color Transparency effects occur as it propagates from the point where it was produced to the point where it undergoes the hard scatter of momentum transfer $\mathbf{q}$ which knocks out the nucleon.  But the large-momentum transfer scattering  at $z_1$ causes the outgoing $\rho$-like configuration to be in a small-sized configuration.  Hence the outgoing $\rho$ experiences reduced interactions on its way out of the nucleus (the knocked-out proton also experiences reduced interactions).  This is a manifestation of Color Transparency effects for small $Q^2$ (but large $t$).  The difference between the CT result and the Glauber result is not as significant, however, as  in the case of large $Q^2$.

\section{Conclusion}
\label{sec:conclusion}

We have calculated the transparency for $\gamma^*+A\to \rho+p+(A-1)^*$, both without inclusion of CT effects (Glauber case) and with inclusion of CT effects, for several different combinations of $A$ and $l_c$.  The transparencies clearly exhibit the coherence length effect, i.e. the decrease of the transparency as $l_c$ is increased, which is not due to Color Transparency.  Thus to observe the effects of CT it is necessary to keep $l_c$ fixed while varying $\nu$ and $Q^2$.  The quantity of experimental interest, namely the integrated transparency, is smaller in general than the transparency evaluated at missing momentum $\mathbf{p}_m=0$.  However, the difference between the Glauber transparency and the CT transparency is marked, particularly as $Q^2$ is increased while $t$ is fixed.  However, it should still be possible to observe the effects of CT when $Q^2$ is small, if $t$ is large enough.  The difference between the CT prediction and the Glauber prediction for the transparency in this case is not as large as it is in the case of large $Q^2$.

\section*{Acknowledgements} This work was partially supported by the DOE grant No. DE-FG02-97ER-41014.

 \end{document}